\documentclass{nature_meth}
\usepackage{amssymb,amsfonts,amsmath}
\usepackage{graphicx} 
\usepackage{multirow}
\usepackage{epsfig}
\usepackage[figoff]{figcaps}
\usepackage{hyperref}
\hypersetup{
    colorlinks=true,
    linkcolor=blue,
    filecolor=magenta,
    urlcolor=cyan,
}
\newcommand{\add}[1] {\textcolor{black}{#1}} 

\usepackage{amssymb}% http://ctan.org/pkg/amssymb
\usepackage{pifont}% http://ctan.org/pkg/pifont
\usepackage{caption,setspace}
\renewcommand{\spacing}[1]{\renewcommand{\baselinestretch}{#1}\large\normalsize}
\spacing{2}
\usepackage{url}
\usepackage{lineno}
\usepackage{booktabs}
\usepackage{adjustbox}
 
\usepackage{marginnote}
\renewcommand{\marginnote}[2][]{}
\usepackage{xr}

\usepackage{graphicx}
\makeatletter
\let\saved@includegraphics\includegraphics
\AtBeginDocument{\let\includegraphics\saved@includegraphics}
\makeatother

\usepackage{kotex}
\usepackage{tablefootnote}
\usepackage{comment}

\title{LLM-driven Multimodal Target Volume Contouring in Radiation Oncology}

\author{Yujin Oh$^{\ast1}$,
Sangjoon Park$^{\ast2,3}$, 
Hwa Kyung Byun$^{4}$, 
{Yeona Cho}$^{5}$, 
{Ik Jae Lee}$^{2}$, 
Jin Sung Kim$^{\dagger2,6}$,
and Jong Chul Ye$^{\dagger7}$
}

\begin{document}
\setstretch{1.2}

\maketitle

\begin{affiliations}
\item{Department of Radiology, Massachusetts General Hospital (MGH) and Harvard Medical School, Boston, MA, USA}
\item{Department of Radiation Oncology, Yonsei University College of Medicine, Seoul, Korea}
\item{Institute for Innovation in Digital Healthcare, Yonsei University, Seoul, Korea}
\item{Department of Radiation Oncology, Yongin Severance Hospital, Yongin, Gyeonggi-do, Korea}
\item{{Department of Radiation Oncology, Gangnam Severance Hospital, Seoul, Korea}}
\item{Oncosoft Inc., Seoul, South Korea}
\item{Kim Jaechul Graduate School of AI, Korea Advanced Institute of Science and Technology (KAIST), Daejeon, Korea}

\item[] $^{\ast}$These authors contributed equally
\item[] $^{\dagger}$Correspondence to Jin Sung Kim (jinsung@yuhs.ac) or Jong Chul Ye (jong.ye@kaist.ac.kr)

\end{affiliations}

\vspace{-0.5cm}
\section*{Abstract}
\begin{abstract}
Target volume contouring for radiation therapy is considered significantly more challenging than the normal organ segmentation tasks as it necessitates the utilization of both image
and text-based clinical information. Inspired by the recent advancement of large language
models (LLMs) that can facilitate the integration of the textural information and images,
here we present an LLM-driven multimodal artificial intelligence (AI), namely LLMSeg, that utilizes the clinical information and is applicable to the challenging task of 3-dimensional context-aware target volume delineation for radiation oncology. We validate our proposed LLMSeg within the context of breast cancer radiotherapy using external validation and data-insufficient environments, which attributes highly conducive to real-world applications. We demonstrate that the proposed multimodal LLMSeg exhibits markedly improved performance compared to conventional unimodal AI models, particularly exhibiting robust generalization performance and data-efficiency.

\end{abstract}

\clearpage
\setstretch{1.6}
%\linenumbers

\section*{Introduction}

Despite the rapid development of Artificial Intelligence (AI) models, there is yet a discernible gap in the realm of medical data processing. Historically, {AI} models have predominantly focused on individual data modalities - either visual or linguistic. This approach starkly contrasts with the intrinsic multimodal practices of physicians, who inherently rely on a confluence of imaging studies and textual electronic medical data for informed decision-making.
 By understanding diverse data types and their interrelationships, multimodal AIs would facilitate more accurate diagnoses, personalized treatment development, and a reduction in medical errors by providing a comprehensive view of patient data. For example, in the field of radiation oncology, which is one of the clinical field to evaluate the potential of multimodal AI applications and the main focus of this article, the integration of multiple modalities {holds great} importance \cite{huynh2020artificial}.

{
 For modern intensity-modulated radiation therapy (IMRT) and its inverse planning, two critical components are needed: organs-at-risk (OARs) and the target volume where the dose is prescribed. OARs are defined as the radiosensitive organs susceptible to damage by ionizing radiation during radiation therapy. Traditionally, they were either manually delineated by human experts or automatically contoured using atlas-based autocontouring algorithms. However, with the advent of deep learning-based AI models, such tasks have been efficiently accomplished \cite{shi2022deep, zhang2023segment}. }
 Therefore, these OARs can be contoured ``as they appear" in the planning computed tomography (CT) images.
 
{However, in contrast to OARs segmentation, the task of target volume delineation, {which also needs to be contoured on the planning CT images but often requires to consider clinical information beyond the visual features, remains crucial for treatment planning and has traditionally been the responsibility of experienced radiation oncologists.} This task is perceived as more challenging due to its intrinsic need for the integration of multimodal knowledge.
Although a multitude of segmentation models have been proposed and explored to enhance the precision and efficacy of this task over the last few years \cite{chung2021clinical, offersen2015estro, choi2020clinical}, a conspicuous gap in research persists, particularly regarding multimodal target delineation \cite{zhang2023segment}.}

This is because the delineation of radiation therapy target transcends beyond the mere consideration of visual elements, such as the gross tumor volume (GTV) \cite{guo2019gross}, and necessitates the incorporation of a myriad of factors, including tumor stage, histological diagnosis, the extent of metastasis, and gene mutation.  
These factors critically influence the potential for occult metastases, which may compromise the survival outcome of a patient. Areas at elevated risk for such metastatic growth are often treated electively, necessitating a clinical 
that is deeply rooted in a comprehensive understanding of various data modalities. Furthermore, additional factors, such as a patient’s performance status and age, which collectively contribute to the general condition, also exert an impact on treatment target delineation. 
Given the imperative nature of considering information beyond imaging in target volume delineation, the application of a multimodal approach in radiation oncology is not merely beneficial but essential for the tasks of the radiation oncology \cite{liu2023artificial}. This is particularly substantiated by the necessity to incorporate textual clinical data, which can significantly influence the identification and subsequent treatment of regions susceptible to occult metastases.

Recently,  large language models (LLMs) -- AI models proficient in processing and generating text, code, and other data types -- have witnessed remarkable advancements \cite{bubeck2023sparks, touvron2023llama, liu2023radiology}. Trained on extensive datasets of text and code, these models discern relationships among varied data types and generate new data, adhering to learned patterns. 
Furthermore, multimodal data such as images, signals, etc., can be easily
integrated into LLMs through adaptors  and generative models for vision understanding and generation, respectively. Consequently, these models have demonstrated promise in a myriad of medical tasks, including multimodal medical report generation, medical question answering, and multimodal segmentation with medical images like chest X-rays \cite{
moor2023foundation, singhal2022large, tu2023towards, lee2023llm}.

Inspired by the multimodal integration capability of LLMs and needs for multimodal information for tumor
target delineation, here we present a 3-dimensional (3D) multimodal clinical target volume (CTV) delineation model, LLMSeg, by integrating clinical information through the LLM for conditioning a segmentation model. 
Specifically, by leveraging the textual information from well-trained LLMs through simple prompt tuning, our cross-attention-based segmentation model has adeptly integrated text-based clinical information into the target volume contouring task. 
More specifically, as illustrated in Fig.~\ref{fig_proposed}(a), we introduce an interactive alignment framework which uses both self-attention and cross-attention mechanisms in a bidirectional manner (text-to-image and image-to-text features), by following the concept of promptable segmentation from Segment Anything Model (SAM) \cite{kirillov2023segment}. To further improve the quality of features, we implement this interactive alignment between all the skip-connected image encoder features with the LLM feature. These layer-wise multimodal features are then combined to jointly predict the target labels through the multimodal decoder. In this way, we ensure the image encoder to efficiently extract meaningful text-related representations and vice versa.
Finally, to transfer the LLM's knowledge while the entire network parameters are kept and achieve superior performance in various downstream tasks~\cite{kim2023zegot, jia2022visual, zhou2022conditional}, we adapt the idea of light-weight learnable text prompts to fully leverage the great linguistic capability of the LLM within the proposed multimodal AI framework.

\add{In this work, we apply LLMSeg to breast cancer target volume delineation task to evaluate its context-aware radiotherapy target delineation performance compared to unimodal AI. Additionally, we expand its application to prostate cancer cases.} By utilizing a well-curated, large-scale dataset from three institutions for development and external validation, we  verify its capability to integrate pivotal \add{clinical information}, such as tumor stage, surgery type, and laterality. 
Experimental results confirm that the model not only demonstrates a significantly enhanced target contouring performance compared to existing unimodal segmentation models but also exhibits behavior that contours targets in accordance with provided clinical information. Notably, the model exhibits superior performance enhancement on an external dataset and shows stable performance gains in data-insufficient settings, demonstrating generalizability and data-efficiency that is not only apt for the characteristics of medical domain data but also align well with the perspective of clinical experts.

\section*{Results}
\paragraph{Accurate and Robust CTV Delineation Performance of Multimodal Model}

Fig.~\ref{fig_proposed}(b) presents a comparative analysis between the vision-only model and our proposed multimodal model for clinical target volume (CTV) delineation in breast cancer patients for all the validation sets. For internal validation, both methods showed promising performance of above 0.8 in the Dice metric, with a substantial improvement is observed in ours. However, the vision-only model showed drastic performance drop of 0.73 and 0.44 in the Dice metric in both external settings. Specifically, in the case of external set \#2, where the manufacturer of acquisition modality differs from that of internal and external set \#1, the vision-only model completely failed to perform CTV delineation. Despite encountering visually shifted data distributions, our multimodal model demonstrated notable stability by consistently maintaining performance across all experimental conditions.

We qualitatively compared two different approaches in Fig.~\ref{fig_proposed}(c). In general, CTV for breast cancer radiation therapy can be categorized into two primary types: one that involves treatment of the breast or chest wall alone, and the other that electively treats the regional lymph nodal area (including axillary, supraclavicular, and internal mammary lymph nodes (LNs)) in addition to the aforementioned areas, given the frequent metastasis of breast cancer to these regions. On the left side of Fig.~\ref{fig_proposed}(c), despite the ground truth label posing CTV on both the breast and regional LNs, the vision-only model only contours the breast alone. Moreover, as the vision-only model lacks information about the laterality of the breast that diagnosed as cancer, partial segmentation masks are observed on the opposite breast. In contrast, the multimodal model accurately contours the breast and regional LNs that need to be treated as CTV. On the right side of Fig.~\ref{fig_proposed}(c), despite early breast cancer case requiring treatment of the breast only, the vision-only model incorrectly includes the regional LN as CTV. Moreover, CTVs are extended to the opposite breast. On the other hands, the multimodal model that integrates the \add{clinical information} accurately contours the requisite treatment areas, encompassing both the breast and the regional LNs, aligning with the ground truth. 

We further compared our methods with other diverse vision-only and multimodal methods in Table~\ref{tab_main}. Our proposed context-aware segmentation, in which the given textual information is not explicitly visible as an actual object in the input image, compared to traditional vision-language segmentation \cite{zhu2024llafs, wang2023visionllm}. Therefore, we adapted publicly available 2D text-driven multimodal segmentation frameworks from various segmentation categories as our baseline models \cite{wang2024hierarchical, lai2023lisa, huemann2023contextual}. Furthermore, we conducted comparisons with two advanced visual backbones \cite{hatamizadeh2022unetr, xing2024segmamba} to justify our selection of the 3D residual U-Net as the visual backbone. In the results shown in Table~\ref{tab_main}, HIPIE \cite{wang2024hierarchical} and LISA \cite{lai2023lisa}, considered SOTA models for 2D referring and reasoning segmentation respectively, showed suboptimal performance in 3D context-aware segmentation. On the other hand ConTEXTualNet \cite{huemann2023contextual}, capable of handling 3D images as inputs, showed promising performance. Nevertheless our approach demonstrated the SOTA performance across all evaluation metrics in various validation settings.

\paragraph{Performance Evaluation by Expert Reveals Superiority of Multimodal Model.}
The assessment of the target volume should not be based on mere metric evaluations such as the Dice, but rather by appropriate clinical rationale. In the context of breast contouring, this involves considerations such as {whether the target volume has been contoured on the treated side of the breast,} the contouring performed on the breast or chest wall depending on the type of surgery (breast-conserving surgery or mastectomy), and whether the regional LNs has been included. Therefore, the appropriateness of target contouring should be evaluated by a board-certified radiation oncologist, ensuring a clinically relevant perspective in the assessment. 
To this end, five rubrics (laterality, surgery type, volume definition, coverage, integrity) were suggested by the board-certified radiation oncologists, to objectively and specifically evaluate the target volume with differentiated scoring reflecting their importance. Detailed descriptions of these rubrics are available in Supplementary Table 1 with Supplementary Fig. 1.

{When evaluated using the proposed rubrics as indicated in Table~\ref{tab_eval}, the multimodal model exhibited superior performance, achieving total scores up to twice as high as those of the vision-only model. Importantly, the model exhibited notably larger gains in rubrics like laterality and volume definition, where incorporation of the clinical context is crucial to achieve accurate results, than in metrics indicative of contouring quality, such as coverage and integrity. This performance gain was particularly pronounced in the external validation, notably in external set \#2, where differences in the image acquisition setting were noted. This demonstrates the multimodal model's robustness and clinical relevance across varied datasets and potential diverse clinical scenarios.}

\paragraph{Data Efficiency and Robustness of the Multimodal Model.}

During the training process of clinical specialists, learning is expedited when textual clinical information is integrated alongside imaging studies, as opposed to focusing on target volume in images alone. This approach facilitates a more rapid assimilation of tendencies and principles of target volume contouring, enabling effective learning even with fewer cases.
We sought to determine whether this efficiency of learning through the integration of textual clinical information could be applied to our multimodal approach.

We observed the performance of each concept in target volume contouring by progressively reducing the size of training dataset. As illustrated in Fig.~\ref{fig_data_efficiency}(a), our multimodal model demonstrated its data efficiency by maintaining stable performance above 0.8 in the Dice even with 40\% of data availability. This starkly contrasts with the vision-only model, whose performance dropped from initial Dice of 0.8 to 0.7. When we utilizing only 20\% of the training dataset, the multimodal model's performance decreased slightly below 0.8 in the Dice, while the vision-only model completely failed to contour CTV in the limited dataset scenario. This performance gap was particularly evident in the external validation results. For external validation \#1, the initial discrepancy between two models was approximately 0.1 in the Dice metric. However, as the size of training dataset decreased, the discrepancy doubled. For external validation \#2,  notable overfitting issues were observed in the vision-only model. On the contrary, our multimodal model achieved robust performance when trained with a reduced dataset of less than 40\%. Qualitative analysis, as depicted in Fig.~\ref{fig_data_efficiency}(b), also supports these results. Detailed quantitative results for all metrics are further provided in Supplementary Table 2.

\paragraph{Differential Target Contouring Based on Varied Textual Inputs.}

To validate the hypothesis that our multimodal model genuinely performs CTV delineation based on textual clinical information, we conducted an experiment to assess whether altering the textual clinical information alone would yield different delineation results, even for the same CT, as illustrated in Fig.~\ref{fig_modify_exp}(a).

As depicted in Fig.~\ref{fig_modify_exp}(b)-(c), the model performed contouring different targets for the same CT, contingent on the provided \add{clinical data}. In Fig.~\ref{fig_modify_exp}(b), for a patient with left breast cancer at stage T1N0M0, upstaging the T stage or N stage demonstrated the inclusion of regional LNs, and altering the tumor's laterality from left to right resulted in contouring on the opposite side. Interestingly, when the type of surgery was changed from breast conserving surgery (BCS) to total mastectomy, it was observed that the previously spared skin was no longer spared, and the target volume was expanded to include the chest wall. For another patient with right breast cancer at stage T2N1M0, as exemplified in Fig.~\ref{fig_modify_exp}(c), downstaging N stage leads to the omission of regional LNs from the designated target volume, and changing the type of surgery to BCS results in a strategic shift to sparing the skin and excluding the pectoralis muscle from the treatment volume. These quantitative results align precisely with the decision policy of radiation oncologists, and substantiate that our model contours the target volume, strongly referencing the textual clinical information as well as the imaging features.

\paragraph{Exploring Textual Clinical Information Provision Methods in the Multimodal Model.}

To demonstrate the necessity of LLM as for our textual clinical information provision method, we conducted an ablation study by replacing our textual module by a simple numeric category method and a CLIP text encoder trained on a relatively smaller textual dataset compared to LLM \cite{radford2021learning}. As indicated in Table~\ref{tab_ablation}(a), the numeric category method, by  representing each clinical information as categorized numbers, exhibited promising performance and showed relatively marginal performance drops to our method in the internal validation setting. However, in the two external validations, the performance gaps were increased up to 0.1 in the Dice metric and significantly more in the HD-95 metric, of up to 10 cm. Moreover, when replacing the textual module with the CLIP ViT-B/16 while maintaining our proposed multiple text prompt tuning method, huge performance gaps were observed compared to our method of up to 0.3 in the Dice metric and up to 10 cm in the HD-95 metric. These findings indicate that the effectiveness of the proposed multimodal model originates from leveraging LLM.

In specific, the numeric category method exhibited the second-most promising performance and showed relatively marginal performance drops to our method in the internal validation setting. However, in the two external validation settings, the performance gaps were increased. Hence, we qualitatively evaluated the source of the performance gap in Fig.~\ref{fig_ablation_omit}(a). In Case \#1, where a patient underwent total mastectomy for T2N1M0 cancer in the left breast, our method accurately contoured the surgically treated breast with an implant, including the regional nodal area in the target volume. However, the numeric category method generated segmentation masks for both breasts, with more mask generation observed on the opposite breast. Similarly, in Case \#2, where a patient underwent breast conservation surgery for T2N1M0 cancer in the left breast, our method accurately included the breast and regional nodes in the target volume while sparing the skin and chest wall. In contrast, the numeric category method only included the breast area in the target volume, excluding the regional nodes, and included parts of the skin and chest wall as like the mastectomy case. Moreover, it partially generated segmentation masks on the opposite breast, demonstrating incomplete reflection of the clinical context.

We further ablated our employment of introducing \add{clinical data} by replacing it with various methodologies. These include utilizing a single or multiple text prompts through prompt tuning, low-rank adaptation (LoRA) fine-tuning \cite{hu2021lora} and directly employing a pre-trained LLM without tuning. As indicated in Table~\ref{tab_ablation}(b), our proposed text prompt tuning method consistently outperformed those using LoRA fine-tuning and no tuning strategy. Moreover, employing multiple learnable text prompts showed an improved performance compared to using a single text prompt. These results indicate that the introduced learnable text prompts were optimized to efficiently fine-tune the LLM for the target volume contouring task.

\paragraph{Ablation Study of Input \add{Clinical Data} Components.}

{We further conducted ablation study by omitting each piece of input \add{clinical information} and compared the difference between a competing method (Numeric Category) and our method (LLMSeg) in Fig.~\ref{fig_ablation_omit}(b)-(c). Firstly, without omission as shown in Fig.~\ref{fig_ablation_omit}(b), our method accurately segmented only the right breast area as the target volume for a person with T1aN0M0 cancer who underwent breast-conserving surgery. 
\add{\marginnote{\add{{R3.C2}}}However, when the information for T stage was removed, the model included some regional nodes in the target range, similar to cases with higher stages.}
This trend was similarly observed in the omission of N stage information, where the model included regional nodes as in cases with nodal metastasis like N1 or N2. 
\add{Likewise, without information about laterality, the model inaccurately contoured the opposite breast.}
On the contrary, the competing model showed inaccurate results such as contouring on the opposite breast even without omission. Moreover, regardless the presence or absence of omission, there was little change in target contouring (e.g., laterality), or target contouring changed in patterns unrelated to the omitted information (e.g., contouring on the opposite side when omitting T stage or N stage information). These results indicate that the competing model receiving clinical context in a simpler manner fail to effectively incorporate such information and perform CTV delineation unrelated to the provided information. Similarly, in another case of T1cN1M0 breast cancer in the left breast where total mastectomy was performed as shown in Fig.~\ref{fig_ablation_omit}(c), 
\add{when surgery information is not provided, our method misidentified the surgery type and produced segmentation results resembling breast-conserving surgery, sparing the skin and chest wall.}
However, the competing model rather contoured on the opposite breast, which was irrelevant to surgical method.}

{In Table~\ref{tab_inputablation}, we further assessed these ablation results quantitatively. For our method, the exclusion of information regarding laterality, which influences the decision on which breast to contour, resulted in the most significant decrease in performance. This was followed by similar degrees of performance decrease upon excluding information related to surgery type and N stage, which impact the inclusion of the skin, chest wall, and the regional nodes. Although excluding T stage information did result in a decrease in performance, it was the least significant, which is rational considering the minimal impact of T stage information on target volume delineation.}

{Overall, these comparative results suggest that the our model considers the clinical context provided in text and is hindered in accurate target volume delineation if any component is missing. That is said, excluding any one component results in lower performance compared to using all available information, indicating that every component contributes to the model's performance.}

\paragraph{Exploring Other Cancer Types.}

{
We further evaluated the proposed multimodal target volume contouring for prostate cancer patients.}
\add{\marginnote{\add{{R3.C3}}}
For prostate cancer, clinical data were directly curated from EMR, as detailed in Supplementary Table 3. This curated EMR data, along with each patient’s age, were then summarized as input clinical data.
}
{Similar to the breast cancer study, we observed the superiority of our multimodal approach over the vision-only approach, with a notable performance gain of up to 0.05 in Dice metric through all the validation settings as shown in Table \ref{tab_prostate}. 
}

{
Similar to breast cancer, an expert evaluation was conducted for prostate cancer. A rubric-based analysis of expert evaluation in Table~\ref{tab_eval_prostate} clearly showed effectiveness of our method. Particularly, these benefits became unequivocally evident in the external validation setting, showing more than double the differences in total scores. Among those necessitating in-depth reference to clinical information for precise scoring — notably, the delineation of the primary site (assessing prostate volume coverage, including the seminal vesicle) and the volume definition (evaluating regional node irradiation appropriateness) — exhibited significantly larger differences compared to the vision-only model. Details on the rubrics used for prostate cancer can be found in the Supplementary Fig. 2 and Supplementary Table 4. 
}

\section*{Discussion}

Despite the promising outcomes demonstrated by AI models in various studies, a notable limitation prevalent in the field of medical AI has been the predominant development of models tailored for singular, specialized tasks \cite{shen2017deep}. For instance, models have been specifically designed and trained to excel in a singular task, such as segmentation \cite{choi2020clinical, chung2021clinical}, diagnosis \cite{de2018clinically, rajpurkar2017chexnet}, or prognosis prediction \cite{choi2019machine, yoo2013osteoporosis}, without the adaptability to transition across various tasks. While these specialized models perform commendably within their designated task, they lack the flexibility to navigate the complex challenges in the medical domain, where the ability to integrate, and concurrently process diverse tasks is crucial.

In the nascent stages of applying vision-language models to the medical domain, initial research endeavors have predominantly focused on the most simple form of vision-text paired data, such as chest radiographs \cite{hosny2018artificial}. These studies have explored various tasks, including zero-shot classification \cite{tiu2022expert}, report generation \cite{moon2022multi, huang2023kiut}, and text-guided segmentation \cite{huemann2023contextual, lee2023llm}. However, the field of radiation oncology emerges as a particularly potent application area for such models \cite{liu2023artificial}.
Radiation oncology exemplifies a robust case for the adoption of multimodality, underpinned by two fundamental factors \cite{huynh2020artificial}. Firstly, decision-making in Radiation Oncology, especially in determining treatment scope and dose, extends beyond imaging to include a plethora of clinical information, such as surgical notes, pathology reports, and electronic medical records, which can be conveyed textually. Secondly, the integration of prior knowledge, including standard treatment guidelines and radiation oncology textbooks, is vital for informed treatment decision-making, with these guidelines also being expressible in textual formats. Consequently, the necessity for multimodality is markedly emphasized in Radiation Oncology (see Supplementary Fig. 3).

Consequently, we have applied LLMs in our research. Our model introduces \add{\marginnote{\add{{R1.C2}}}several aspects with substantial clinical value} and has demonstrated commendable results by accurately segmenting radiation therapy target volume based on clinical information, thereby achieving absolute performance where the multimodal model surpasses the vision-only model. It also exhibits a pronounced performance differential in external validation settings and demonstrates data-efficiency in data-insufficient settings. This resonates intriguingly with the clinical implications, especially mirroring the learning trajectory and characteristics of clinical experts. In the clinical training of experts, reliance is placed on multimodality information; learning is not confined to either images or text but is rather a confluence of both, facilitating the inference of text-image relationships and enabling effective learning even with relatively fewer cases. This aspect of the clinical learning paradigm, being data-efficient, aligns seamlessly with our proposed multimodal model.

The decrement in classical AI-driven delineation generalization performance is often attributed to variations in image acquisition settings and characteristics of devices from different vendors, among other factors. 
Nonetheless, the ability of clinical experts to perform target contouring is scarcely influenced by external factors such as CT scanning conditions.
This is because linguistical concepts embodied in textual clinical information, are independent of such acquisition settings. Therefore, it is plausible that our model, which learns in conjunction with such textual clinical information by leveraging the great linguistic capability of LLMs, demonstrates particularly commendable performance in external validation settings. This characteristic is particularly optimal for the medical domain, where training data is often limited and stable generalization performance is a prerequisite across varied external settings, thereby heralding a promising future for the application of multimodal models in medical AI.

Furthermore, we have demonstrated the necessity of incorporating \add{clinical information} into target volume contouring, particularly in cases such like breast cancer where the GTV may not be clearly visible in the planning CT image. This necessity is highlighted through diverse comprehensive qualitative comparison. In Fig.~\ref{fig_proposed}(c), where the inclusion of clinical context is crucial for for both cases, the multimodal target contouring reflects comprehensive considerations of clinical context. This necessity becomes more evident, where the absence of clinical context in the vision-only model results in clear failure cases. Additionally, in the detailed rubric comparison between the vision-only model and our multimodal model presented in Table~\ref{tab_eval} and Table~\ref{tab_eval_prostate}, the largest gains are observed in metrics that can be achieved through clinical considerations, such as laterality and volume definition. These results further emphasize the value of our multimodal approach.

Our study has several limitations.  First, our evaluation is confined to patients at their initial diagnosis, leaving a scope for further exploration into varied patient scenarios and treatment stages, which can potentially influence the model's applicability and performance. Second, the model does not incorporate considerations for radiation therapy doses in target volume contouring, presenting an opportunity to explore how dose-related variables could be integrated to enhance delineation and treatment planning in future studies. Third, while the model utilizes refined, rather than raw, clinical data, future research can explore mechanisms for automating the data refinement process or further develop capabilities to process raw clinical data, thereby reducing the need for manual intervention and potentially uncovering additional insights from unstructured clinical reports.
Fourth, although our research scope covers both breast and prostate cancers to confirm its applicability in various cancer types, these cancer types are categorized as having relatively standardized target volume. This suggests the necessity for further validation of our method's generalizability across a wider range of cancer types, which demand more challenging and intricate clinical considerations for accurate target volume delineation.
Fifth, in our work, we focuses CTV contouring to clearly demonstrate advantages of our multimodal model. However, GTV delineation, which involves contouring visually apparent areas, is crucial in clinical practice due to its importance in boost techniques for increased dose administration in many cancer types.

Additionally, in cancer types where the target volume is primarily determined based on GTV, such as lung cancer \cite{hosny2022clinical}, the benefits of integrating clinical information through our method may be relatively limited. Therefore, it is necessary to validate whether our method still offers utility in such cancer types, where the emphasis is on GTV for target volume definition.
{
Therefore, future studies should expand to encompass GTV contouring, thereby improving its clinical utility.
}
Last, but not least, the black-box nature of AI may hinder clinician's direct utilization. Therefore, our proposed model should provide explainable results such like confidence map in the clinical practice, as shown in Supplementary Fig. 4. These visual clues enable clinicians to interpret the model output by referencing the level of confidence for each segment of contour.

Despite aforementioned limitations, our research serves as a pivotal step towards the multimodal models in the field of radiation oncology, verifying the clinical utility and emphasizing the significance of intertwining textual clinical data with medical imaging. The model proposes a pathway for crafting more adaptable and clinically pertinent AI models in medical imaging and treatment planning. Future research would refine and broaden such models, closer to harnessing the full potential of multimodal framework in elevating clinical decision-making and patient care.

\section*{Methods}
\label{method}

\subsection{Ethic Committee Approval.}
The hospital data deliberately collected for this study were ethically approved by the Institutional Review Board of Department of Radiation Oncology at Yonsei Cancer Center, Department of Radiation Oncology at Yongin Severance Hospital, and Department of Radiation Oncology at Gangnam Severance Hospital (approval numbers of 4-2023-0179, 9-2023-0161 and 3-2023-0396 for each). The requirement for informed consent was waived due to the retrospective nature of the study.

% \subsection{Definition of Task}

% {
% In radiation oncology, accurately delineating treatment target volumes is crucial, involving three key categories: Gross Tumor Volume (GTV), Clinical Target Volume (CTV), and Planning Target Volume (PTV). GTV focuses on the visible tumor, while CTV, occasionally derived directly from GTV, often also includes regions prone to microscopic disease, incorporating various clinical factors, such as pathology, stage, and other tumor and patient related factors.
% }

% PTV further includes margins to accommodate uncertainties in patient positioning. Specifically, in breast cancer treatment, target volumes vary significantly depending on the stage of the disease and the presence of lymph node metastasis, underlining the complexity of radiation therapy planning beyond mere image segmentation. To address this, we aimed to develop a model that incorporates clinical information—mimicking an experienced oncologist’s decision-making process—in delineating target volumes for breast cancer patients at their initial diagnosis. Further details of the definition of task and target volume are provided in the Supplementary Section~\ref{supple_task} and Supplementary Section~\ref{supple_ctv}.

\subsection{Schematic Comparison of the Workflows of Radiology and Radiation Oncology.}
Supplementary Fig. 3 delineates the clinical workflows in Radiology and Radiation Oncology. {
In radiology, while the patient's history, previous diagnoses, past treatments, and previous imaging results are comprehensively considered, the most crucial element remains the findings visible in the current images, thus heavily relying on the visual information of the current imaging study. Conversely, in radiation oncology, determining the treatment target volume and prescribing doses necessitates a more comprehensive consideration of the patient's history, pre- and post-operative imaging results, surgical pathology findings, laboratory results, and other clinical information, resulting in a relatively less reliance on the current simulation CT images.
}

Additionally, the integration of prior knowledge, including standard treatment guidelines and radiation oncology textbooks, is crucial for informed treatment decision-making and can also be expressed in textual formats. Therefore, the significance of multimodal apporoach is notably enhanced in Radiation Oncology compared to Radiology.

\subsection{Definition of Task}
\label{supple_task}

{
In radiation oncology, the treatment target volumes are categorized into Gross Tumor Volume (GTV), Clinical Target Volume (CTV), and Planning Target Volume (PTV). GTV corresponds to the visible tumor, align with traditional segmentation's objective to delineate visible image portions.} 
{CTV, while occasionally derived directly from GTV in the presence of a gross tumor, often also includes regions prone to microscopic disease. This necessitates the incorporation of diverse clinical factors, such as tumor type, histological findings, cancer stage (TNM classification), patient age, and performance status in specific cases. PTV further expands upon CTV to include margins that account for uncertainties in patient setup and positioning. Consequently, achieving accurate target volume delineation in radiation oncology goes beyond the scope of traditional segmentation task, necessitating incorporation of a various clinical context as well as the structures visible on the CT scan.

Taking breast cancer as an example, {in early-stage cases (e.g., stage I) where there is no regional lymph nodes (LN) metastasis, often only the whole breast is included in the radiation therapy target volume. On the other hand, in advanced stages (e.g., stage IIIB), where regional LN metastasis is identified during surgery, there is often a need for elective nodal irradiation across all regional nodal areas.} However, such distinctions are not discernible during the CT simulation for post-operative radiation therapy planning and require acquisition through other forms of information.
Consequently, we aimed to develop a model that can consider clinical information such as primary tumor type, stage, age, and performance status in a manner akin to an experienced radiation oncologist by providing such data in the form of textual information to a multimodal model.

Among the primary cancer types, we initially targeted breast cancer. This was predicated on the fact that breast cancer presents with relatively uniform guidelines for target delineation according to the clinical information including primary tumor location, size, and the presence of nodal metastasis, etc. Furthermore, the inter-observer variability in target delineation for breast cancer is also expected to be small compared with other cancer types.
Within the task of radiation therapy target delineation for breast cancer, we exclusively incorporated cases of patients at their initial diagnosis of breast cancer. This decision was based on the understanding that treatments with aims such as salvage or palliative often exhibit significant variability according to the preferences of the physicians as well as the patients, and other circumstances.

\subsection{Details of Clinical Target Volume.}
\label{supple_ctv}

{ 
For breast cancer, the CTV for early breast cancer (Tis-T2) without nodal metastasis at initial diagnosis is limited to the whole breast. For those with nodal metastasis or in cases of locally advanced breast cancer (T3-4), as well as T2 cases with adverse features without proper axillary dissection, regional node irradiation was primarily considered. The delineation of regional nodes, especially the level of inclusion for the supraclavicular lymph node, is defined according to the Radiation Therapy Oncology Group (RTOG) guidelines for cases identified with N2 or more nodal metastasis, and by the European Society for Radiotherapy and Oncology (ESTRO) guidelines for instances with N1 or less nodal involvement.
}

{
For prostate cancer, the definition of the CTV involved a more complex consideration of factors. In the presence of pelvic LN, regional node irradiation was performed in conjunction with prostate bed radiation. The decision to perform elective nodal irradiation on the pelvic LNs was based on the National Comprehensive Cancer Network risk groups, taking into account a combination of factors such as T stage, Prostate-Specific Antigen (PSA) levels, and Gleason score, particularly for those classified within the very high and high-risk groups. However, in individuals aged 80 and over, consideration of age led to the omission of pelvic LN irradiation. In cases where pathologic or imaging findings confirmed seminal vesicle invasion, contouring was performed to include the prostate and extend to the seminal vesicles within the CTV.
}

\subsection{Details of Datasets.}
For model development and internal validation, we acquired data from {981} patients treated at the Department of Radiation Oncology at Yonsei Cancer Center between September 2021 and October 2023. These patients had been initially diagnosed with the breast cancer and underwent radiation therapy post-curative surgery with the primary objective of preventing recurrence. To better reflect real clinical application, 
the ideal approach for external validation needs the use of patient data acquired under different conditions and with equipment from a different vendor. Therefore, we utilized data from {206} patients treated at the Department of Radiation Oncology at Yongin Severance Hospital. {We further utilized data from 204 patients treated at the Department of Radiation Oncology at Gangnam Severance Hospital.} We confirmed that the external cohort was non-overlapping with those included in the model development nor internal validation.

{Supplementary Table 5 presents characteristics of breast cancer patients for each dataset.} Across the train, internal, and external validation sets, distributions of factors such as location and T stage were observed to be consistent. The proportion of patients with LN metastasis and those undergoing total mastectomy was higher in the train and internal validation sets than the external validation set. 
{Furthermore, due to the more advanced stages of disease, the proportion of patients who underwent neoadjuvant chemotherapy prior to surgery was observed to be higher in the train and internal validation sets compared to the external validation sets.} Consequently, the percentage of patients receiving irradiation to the chest wall and regional LNs was also higher in the train and internal validation sets compared to the external validation set.
{
When compared to the training and internal validation sets, external set \#1 exhibited similar imaging equipment and conditions. However, external set \#2 presented differences in image acquisition conditions, such as vendor, filter type, and slice thickness.
}

For evaluating the proposed method for other cancer types, we further acquired data from 943 prostate cancer patients from Yonsei Cancer Center and 141 prostate cancer patients from Yongin Severance Hospital. We confirmed that the external cohort was non-overlapping with those included in the model development nor internal validation. 
{Supplementary Table 6 presents characteristics of prostate cancer patients for each dataset.}
{
In terms of the distribution of T and N stages, as well as Gleason scores, the training, internal validation, and external validation sets demonstrated a relatively uniform distribution. However, the initial PSA levels were found to be higher in the training and internal validation sets compared to the external validation set. Additionally, the proportion of individuals undergoing prostatectomy was also higher in the training and internal validation sets, which consequently led to a higher percentage of patients receiving radiotherapy with a definitive aim in the external validation set, while those in the training and internal validation sets were more likely to receive radiotherapy with a salvage aim. There were no significant differences in image acquisition settings across the datasets.
}

We not only utilized patient's simulation CT images and CTVs for radiation therapy, but also incorporated text-based clinical information that is essential for precise target delineation. This additional information included the location of the primary cancer, type of surgery undertaken, disease stage, and the status of nodal metastasis. 
{The input \add{clinical data} was prepared by the tabular format derived from raw clinical data for breast cancer, as shown in Supplementary Table 3(a). The resulting clinical context was then curated using custom criteria. Initially, these criteria were devised by a board-certified radiation oncologist. Subsequent refinement was achieved through ablation studies on the components to construct the most effective \add{clinical information}, and the resulting examples of input texts are illustrated in the right-most column. }

{Compared to breast cancer, by utilizing tabular structure of clinical data which is curated by clinicians, for prostate cancer, we directly curated input clinical information from EMR data, by utilizing 10-shot in-context learning strategy with a pre-trained LLM, as shown in Supplementary Table 3(b). Then, the curated EMR Data and each patient's age were summarized as input \add{clinical data} in the right-most column. In the future study, the similar in-context learning approach can be applied to breast cancer study for an automated framework. 
}

\subsection{Details of Implementation.}
The schematic of our multimodal AI is illustrated in Fig.~\ref{fig_proposed}. For the image encoder/decoder and the large language model (LLM), we employed the 3D Residual U-Net \cite{cciccek20163d} and the pre-trained Llama2-7B-chat \cite{touvron2023llama} model, respectively. For the interactive alignment modules, we utilized the two-way transformer modules of SAM \cite{kirillov2023segment}.  %Details of multimodal framework are deferred to Supplementary Section~\ref{supple_multimodal}. {Details of the network implementation and training complexity is further specified in Supplementary Section~\ref{supple_implement}.}
We further propose detailed multimodal AI framework  as illustrated in Supplementary Fig. 5. We introduce three key components: (a) text prompt tuning, (b) multimodal interactive alignment, and (c) CTV delineation.

\subsection{(a) Text Prompt Tuning} To efficiently fine-tune the large language model (LLM), we introduce $N$-text prompts $\mathcal{V} = \{{v^{n}|^N_{n=1}}\}$ as illustrated in Supplementary Fig. 5(a), where each $v^n \in \mathbb{R}^{M \times D}$ consists of $M$ vectors with the dimension $D$, which is same embedding dimension as the LLM. These learnable vectors are randomly initialized, and then consistently prepended to each of tokenized {clinical data}, which denoted as  $[$TEXT$]$ tokens.  We additionally append a token, denoted as [SEG], which is intended to attend to all the aforementioned vectors and tokens. Here, the final prompted text input $t$ can be formulated as follows: 
\begin{align}
t = \{v_1^n, v_2^n, ..., v_M^n, [\text{TEXT}], [\text{SEG}]\}.
 \label{eq:textembedding}
\end{align}
Then, using the prompted text input $t$, the frozen LLM results the context embeddings ${g} \in {\mathbb{R}^{N \times D}}$ as output embeddings as for the inputted [SEG] token.

\subsection{(b) Multimodal Interactive Alignment } 
To align the context embeddings ${g}$ with the image embeddings ${f_l} \in \mathbb{R}^{H_l W_l S_l \times C_l}$, where $f_l$ is the $l$-th layer output of the 3D image encoder, $H_{l}$, $W_{l}$, and $S_{l}$ correspond to height, width, and slice of the image embeddings, and $C_{l}$ is the intermediate channel dimension of each $l$-th layer output, we first project ${g}$ to have the identical dimension with that of each ${f_l}$ through layer-wise linear layer. As illustrated in Supplementary Fig. 5(b), the linearly projected context embeddings ${\bar{g}}_l$ are then self-attended and crossly-attended with the image embedding ${f_l}$ to result context-aligned image embeddings ${f_l}^*$. Detailed specifications of each $l$-th layer embeddings and the interactive alignment module are listed in Supplementary Table 7.

\subsection{(c) CTV Delineation } 
After the multimodal interactive alignment, the context-aligned image embeddings ${f_l}^*$ become inputs for the 3D image decoder. As illustrated in Supplementary Fig. 5(c), for the final predicted output $\hat{y}$, we calculated the combination of the Cross-entropy (CE) loss and the Dice coefficient (Dice) loss by following: 
\begin{align}
\min_{\mathcal{D},\mathcal{V}} \mathcal{L} = \lambda_{\text{ce}} \mathcal{L}_{\text{ce}}(\hat{y}, y)  + \lambda_{\text{dice}}\mathcal{L}_{\text{dice}}(\hat{y}, y) , \nonumber\\
\text{where}  \; \mathcal{L}(\hat{y}, y) = -\mathbb{E}_{x\sim P_X} \left[y_i \log p(\hat{y}_i)\right],
\label{losses}
\end{align}
\noindent where $\mathcal{D}$ denotes our prospoed LLMSeg, $\mathcal{V}$ denotes multiple text prompts, $\lambda_{\text{ce}}$ and $\lambda_{\text{dice}}$ are hyper-parameters for each CE loss and Dice loss, respectively. $y \in \mathbb{R}^{B \times H W S}$ is the 3D ground-truth CTV mask, where $B$ denotes batch size, $H$, $W$, and $S$ correspond to height, width, and slice of ground-truth CTV mask.  $p(\hat{y}_i)$ denotes softmax probability of the i-th pixel within the final predicted output $\hat{y} \in \mathbb{R}^{B \times H W S}$, which is denfied as:
\begin{align}
\hat{y} = \mathcal{D}(x, t)\, 
\label{network_output}
\end{align}
\noindent where $x \in \mathbb{R}^{B \times H W S}$ is input 3D CT scan, $t$ is prompted clinical data corresponds to the input CT scan $x$ with text prompts $\mathcal{V}$.  

%\subsection{(d) Hyper-parameter Settings } 
 % by varying experimental conditions.

\subsection{Details of Network Training.}
\label{supple_implement}

When pre-processing the data, all the chest CT images and CTVs were initially re-sampled to have an identical voxel spacing of 1.0 $\times$ 1.0 $\times$ 3.0 $mm^3$. The image intensity values were truncated between -1,000 and 1,000 of Hounsfield unit (HU), and linearly normalized within a range between 0 and 1.0.
When training the network, a 3D patch with a size of 384 $\times$ 384 $\times$ 128 pixels was randomly cropped to cover the entire breast alongside with its paired {clinical data} with batch size of 2. When evaluating the trained network, the entire 3D CT image was tested using sliding windows with a 3D patch with a size of 384 $\times$ 384 $\times$ 128 pixels. We set the optimal hyper-parameters as listed in Supplementary Table 8. During training, we let the entire LLM frozen, while made the image encoder/decoder modules, the interactive alignment modules and their corresponding linear layers, and the text prompts as trainable parameters

As the loss function, we computed both the binary cross-entropy loss and the Dice loss, with the weight value for each loss as 1.0, respectively. The network parameters were optimized using AdamW \cite{KingBa15} optimizer with a learning rate of 0.0001, until the training epoch reaching 100. We implemented the network using the open-source library MONAI\footnote{\url{https://monai.io/}}. All the experiments were conducted using the PyTorch \cite{paszke2019pytorch} in Python using CUDA 11.4 on NVIDIA RTX A6000 48GB. We further described backbones for each model, and compared training complexity in Supplementary Table 9.

\subsection{Rationales of Selecting Baseline Models.}

Our baselines, ConTEXTualNet\cite{huemann2023contextual}, LISA\cite{lai2023lisa}, and  HIPIE\cite{wang2024hierarchical} along with our proposed model, LLMSeg, are designed to extract characteristics from an input sentence that are not explicitly visible in the image, as categorized in Supplementary Table 10. For example, tasks may include identifying the food item richest in Vitamin C from an image and generating a segmentation mask, or recognizing medical conditions and treatment plans (like cT2, N1mi, breast conserving surgery, and left-side procedures). These tasks necessitate a deep understanding of the sentence context and the ability to infer answers for context-aware or reasoning/referring-based segmentation. Both ConTEXTualNet, LISA and HIPIE, like our model, leverage text embeddings derived from a language model to facilitate multimodal segmentation.

Additionally, for a meaningful comparative study, it is crucial to retrain the baseline models with our 3D CT training data. ConTEXTualNet, being a CNN-based network designed for end-to-end training, allows us to adapt the original 2D model into a 3D model suitable for retraining with our 3D data.  On the other hand, recent SOTA multimodal foundation models for segmentation, such as LISA\cite{xu2023open} and HIPIE\cite{wang2024hierarchical}, utilize 2D SAM\cite{kirillov2023segment} or CLIP\cite{radford2021learning}-based cross-attention modules. Adapting these models to process 3D volumes as a whole would require retraining the 2D foundation model with 3D data, which is not feasible given our constraints. Consequently, to preserve their transfer learning mechanism based on the frozen 2D foundation model, we retrain these models by converting 3D CT scans to 2D slices as inputs. This highlights a limitation of current 2D vision-language models when adapting to 3D images, resulting in the loss of volumetric context for clinical information-guided multimodal segmentation and yielding suboptimal performance.

The reason for not including traditional open-vocabulary segmentation models in our study is that they are designed for semantic segmentation of visually discernible objects in an image, such as walls, chairs, windows, floors, and ceilings, as depicted in Supplementary Table 10. This capability stems from their use of pre-trained 2D vision-language foundation models which serves as their frozen backbone for feature extraction. These models leverage pre-aligned word-image features for semantic segmentation, thus, there are not an appropriate baselines for our medical context-aware segmentation purposes, as the radiotheraphy target volumes in CT images are not visually identifiable.

\subsection{Details of Evaluation.} {To quantitatively evaluate the CTV delineation performance, we calculated Dice coefficient (Dice), Intersection over Union (IoU), and the 95th percentile of Hausdorff Distance (95-HD) \cite{crum2006generalized} to measure spatial distances between the ground-truth and the predicted contours. When calculating the 95-HD, all the measured distances in the pixel unit are converted with respect to the original pixel resolution, and the results are expressed in centimeters (cm).}

\subsection{{Details of Clinical Evaluation.}}

To accurately assess the performance of the model, we conducted clinical evaluations by the board-certified radiation oncologist with over five years of experience. To provide a more detailed evaluation of the model's performance and establish an objective criterion for assessment, we employed rubrics proposed by the radiation oncologists. For breast cancer, these rubrics included laterality (right, left, or bilateral - 1 point), type of surgery (whether the case was post-breast conserving surgery (BCS) or mastectomy - 1 point), volume definition (accurate definition of breast or chest wall, inclusion of regional lymph nodes - 1.5 points), coverage (ensuring the target volume was adequately covered without encompassing unnecessary areas), and integrity (absence of incomplete or distorted segmentation output), constituting a total of 5 points. Detailed criteria for each rubric and illustrative examples are provided in Supplementary Fig. 1 and Supplementary Table 1.

For prostate cancer, the criteria included primary site (accuracy in defining the treatment scope for the prostate, including seminal vesicles), volume definition (appropriate inclusion of the prostate and regional nodes), coverage, and integrity, totaling 4 points. The rubrics of laterality, surgery type, volume definition, and primary site were established to assess the appropriateness of the underlying concepts in defining the scope of the target area. Conversely, the criteria for coverage and integrity were specifically designed to evaluate the quality of the contouring. Detailed criteria for each rubric and illustrative examples are provided in Supplementary Fig. 2 and Supplementary Table. 4. 

Utilizing these evaluation criteria, to ensure fairness, the same board-certified radiation oncologists conducted assessments of the segmentation outputs by comparing them with the ground truth and considering the clinical context, all while being blinded to whether the outputs were generated by a vision-only model or a multimodal model.

%\subsection{{Visualization of Confidence Level.}}
%\label{supple_confidence}
%{We additionally provide confidence level for our segmentation results in the practical clinical scenario as shown in Supplementary Fig. 4, enabling clinicians to interpret the model output by referencing the level of confidence for each segment of contour. } 

\subsection{Statistics \& Reproducibility.} 
{For statistical analysis, we used the non-parametric bootstrap method to calculate the confidence interval (CI) for each metric. We randomly sampled the total size of dataset from the original dataset while allowing replacement for 1,000 times, repeatedly. Then, the mean values and the 95th percentile of confidence intervals were estimated from the relative frequency distribution of each trial. Two-tailed Student’s paired t-test was used for the statistical comparison between the two groups. No statistical method was used to predetermine sample size. No data were excluded from the analyses; The experiments were not randomized; The Investigator was not blinded to allocation during experiments and outcome assessment.}

\section*{Data Availability}
Individual patient data, including raw clinical and imaging data, cannot be readily shared due to the protection of individual privacy. Data sharing of de-identified clinical and imaging data used in this study may be considered upon contacting the corresponding author, J.C.Y. (jong.ye@kaist.ac.kr), but will require review and approval by each institution’s IRB. Initial requests will receive a response within 10 working days. Depending on the specific circumstances of the country and institution, data sharing may not be possible, and if approved, the process may take between 2 months to 1 year.
We have made available all numerical data, as well as structured clinical information for cases with qualitative results, used to generate the figures and tables in this study. These data are included in the source data provided with this work. A small sample of subjects with similar characteristics has also been made available for validation purposes at {https://github.com/tvseg/MM-LLM-RO}\cite{tvseg_2024_12792278}. No additional documents, such as study protocols or statistical analysis plans, will be provided. While individual patient data will not be directly shared, the open-source sample data will remain available indefinitely. Data usage is restricted to research purposes only, and redistribution is prohibited.

\section*{Code Availability}
The Pytorch codes for the  proposed Multimodal AI used in this study is available at the following Github repository at {https://github.com/tvseg/MM-LLM-RO}\cite{tvseg_2024_12792278}.

\section*{References}
\bibliographystyle{naturemag}
\bibliography{refs}

\begin{thebibliography}{10}
\expandafter\ifx\csname url\endcsname\relax
  \def\url#1{\texttt{#1}}\fi
\expandafter\ifx\csname urlprefix\endcsname\relax\def\urlprefix{URL }\fi
\providecommand{\bibinfo}[2]{#2}
\providecommand{\eprint}[2][]{\url{#2}}

\bibitem{huynh2020artificial}
\bibinfo{author}{Huynh, E.} \emph{et~al.}
\newblock \bibinfo{title}{Artificial intelligence in radiation oncology}.
\newblock \emph{\bibinfo{journal}{Nature Reviews Clinical Oncology}}
  \textbf{\bibinfo{volume}{17}}, \bibinfo{pages}{771--781}
  (\bibinfo{year}{2020}).

\bibitem{shi2022deep}
\bibinfo{author}{Shi, F.} \emph{et~al.}
\newblock \bibinfo{title}{Deep learning empowered volume delineation of
  whole-body organs-at-risk for accelerated radiotherapy}.
\newblock \emph{\bibinfo{journal}{Nature Communications}}
  \textbf{\bibinfo{volume}{13}}, \bibinfo{pages}{6566} (\bibinfo{year}{2022}).

\bibitem{zhang2023segment}
\bibinfo{author}{Zhang, L.} \emph{et~al.}
\newblock \bibinfo{title}{Segment anything model (sam) for radiation oncology}.
\newblock \emph{\bibinfo{journal}{arXiv preprint arXiv:2306.11730}}
  (\bibinfo{year}{2023}).

\bibitem{chung2021clinical}
\bibinfo{author}{Chung, S.~Y.} \emph{et~al.}
\newblock \bibinfo{title}{Clinical feasibility of deep learning-based
  auto-segmentation of target volumes and organs-at-risk in breast cancer
  patients after breast-conserving surgery}.
\newblock \emph{\bibinfo{journal}{Radiation Oncology}}
  \textbf{\bibinfo{volume}{16}}, \bibinfo{pages}{1--10} (\bibinfo{year}{2021}).

\bibitem{offersen2015estro}
\bibinfo{author}{Offersen, B.~V.} \emph{et~al.}
\newblock \bibinfo{title}{Estro consensus guideline on target volume
  delineation for elective radiation therapy of early stage breast cancer}.
\newblock \emph{\bibinfo{journal}{Radiotherapy and oncology}}
  \textbf{\bibinfo{volume}{114}}, \bibinfo{pages}{3--10}
  (\bibinfo{year}{2015}).

\bibitem{choi2020clinical}
\bibinfo{author}{Choi, M.~S.} \emph{et~al.}
\newblock \bibinfo{title}{Clinical evaluation of atlas-and deep learning-based
  automatic segmentation of multiple organs and clinical target volumes for
  breast cancer}.
\newblock \emph{\bibinfo{journal}{Radiotherapy and Oncology}}
  \textbf{\bibinfo{volume}{153}}, \bibinfo{pages}{139--145}
  (\bibinfo{year}{2020}).

\bibitem{guo2019gross}
\bibinfo{author}{Guo, Z.}, \bibinfo{author}{Guo, N.}, \bibinfo{author}{Gong,
  K.}, \bibinfo{author}{Li, Q.} \emph{et~al.}
\newblock \bibinfo{title}{Gross tumor volume segmentation for head and neck
  cancer radiotherapy using deep dense multi-modality network}.
\newblock \emph{\bibinfo{journal}{Physics in Medicine \& Biology}}
  \textbf{\bibinfo{volume}{64}}, \bibinfo{pages}{205015}
  (\bibinfo{year}{2019}).

\bibitem{liu2023artificial}
\bibinfo{author}{Liu, C.} \emph{et~al.}
\newblock \bibinfo{title}{Artificial general intelligence for radiation
  oncology} (\bibinfo{year}{2023}).
\newblock \eprint{2309.02590}.

\bibitem{bubeck2023sparks}
\bibinfo{author}{Bubeck, S.} \emph{et~al.}
\newblock \bibinfo{title}{Sparks of artificial general intelligence: Early
  experiments with gpt-4}.
\newblock \emph{\bibinfo{journal}{arXiv preprint arXiv:2303.12712}}
  (\bibinfo{year}{2023}).

\bibitem{touvron2023llama}
\bibinfo{author}{Touvron, H.} \emph{et~al.}
\newblock \bibinfo{title}{Llama 2: Open foundation and fine-tuned chat models}.
\newblock \emph{\bibinfo{journal}{arXiv preprint arXiv:2307.09288}}
  (\bibinfo{year}{2023}).

\bibitem{liu2023radiology}
\bibinfo{author}{Liu, Z.} \emph{et~al.}
\newblock \bibinfo{title}{Radiology-gpt: A large language model for radiology}.
\newblock \emph{\bibinfo{journal}{arXiv preprint arXiv:2306.08666}}
  (\bibinfo{year}{2023}).

\bibitem{moor2023foundation}
\bibinfo{author}{Moor, M.} \emph{et~al.}
\newblock \bibinfo{title}{Foundation models for generalist medical artificial
  intelligence}.
\newblock \emph{\bibinfo{journal}{Nature}} \textbf{\bibinfo{volume}{616}},
  \bibinfo{pages}{259--265} (\bibinfo{year}{2023}).

\bibitem{singhal2022large}
\bibinfo{author}{Singhal, K.} \emph{et~al.}
\newblock \bibinfo{title}{Large language models encode clinical knowledge}.
\newblock \emph{\bibinfo{journal}{arXiv preprint arXiv:2212.13138}}
  (\bibinfo{year}{2022}).

\bibitem{tu2023towards}
\bibinfo{author}{Tu, T.} \emph{et~al.}
\newblock \bibinfo{title}{Towards generalist biomedical ai}.
\newblock \emph{\bibinfo{journal}{arXiv preprint arXiv:2307.14334}}
  (\bibinfo{year}{2023}).

\bibitem{lee2023llm}
\bibinfo{author}{Lee, S.}, \bibinfo{author}{Kim, W.~J.},
  \bibinfo{author}{Chang, J.} \& \bibinfo{author}{Ye, J.~C.}
\newblock \bibinfo{title}{Llm-cxr: Instruction-finetuned llm for cxr image
  understanding and generation} (\bibinfo{year}{2024}).
\newblock \eprint{2305.11490}.

\bibitem{kirillov2023segment}
\bibinfo{author}{Kirillov, A.} \emph{et~al.}
\newblock \bibinfo{title}{Segment anything}.
\newblock \emph{\bibinfo{journal}{arXiv preprint arXiv:2304.02643}}
  (\bibinfo{year}{2023}).

\bibitem{kim2023zegot}
\bibinfo{author}{Kim, K.}, \bibinfo{author}{Oh, Y.} \& \bibinfo{author}{Ye,
  J.~C.}
\newblock \bibinfo{title}{Zegot: Zero-shot segmentation through optimal
  transport of text prompts}.
\newblock \emph{\bibinfo{journal}{arXiv preprint arXiv:2301.12171}}
  (\bibinfo{year}{2023}).

\bibitem{jia2022visual}
\bibinfo{author}{Jia, M.} \emph{et~al.}
\newblock \bibinfo{title}{Visual prompt tuning}.
\newblock \emph{\bibinfo{journal}{arXiv preprint arXiv:2203.12119}}
  (\bibinfo{year}{2022}).

\bibitem{zhou2022conditional}
\bibinfo{author}{Zhou, K.}, \bibinfo{author}{Yang, J.}, \bibinfo{author}{Loy,
  C.~C.} \& \bibinfo{author}{Liu, Z.}
\newblock \bibinfo{title}{Conditional prompt learning for vision-language
  models} (\bibinfo{year}{2022}).
\newblock \eprint{2203.05557}.

\bibitem{zhu2024llafs}
\bibinfo{author}{Zhu, L.}, \bibinfo{author}{Chen, T.}, \bibinfo{author}{Ji,
  D.}, \bibinfo{author}{Ye, J.} \& \bibinfo{author}{Liu, J.}
\newblock \bibinfo{title}{Llafs: When large language models meet few-shot
  segmentation} (\bibinfo{year}{2024}).
\newblock \eprint{2311.16926}.

\bibitem{wang2023visionllm}
\bibinfo{author}{Wang, W.} \emph{et~al.}
\newblock \bibinfo{title}{Visionllm: Large language model is also an open-ended
  decoder for vision-centric tasks} (\bibinfo{year}{2023}).
\newblock \eprint{2305.11175}.

\bibitem{wang2024hierarchical}
\bibinfo{author}{Wang, X.} \emph{et~al.}
\newblock \bibinfo{title}{Hierarchical open-vocabulary universal image
  segmentation}.
\newblock \emph{\bibinfo{journal}{Advances in Neural Information Processing
  Systems}} \textbf{\bibinfo{volume}{36}} (\bibinfo{year}{2024}).

\bibitem{lai2023lisa}
\bibinfo{author}{Lai, X.} \emph{et~al.}
\newblock \bibinfo{title}{Lisa: Reasoning segmentation via large language
  model} (\bibinfo{year}{2023}).
\newblock \eprint{2308.00692}.

\bibitem{huemann2023contextual}
\bibinfo{author}{Huemann, Z.}, \bibinfo{author}{Hu, J.} \&
  \bibinfo{author}{Bradshaw, T.}
\newblock \bibinfo{title}{Contextual net: A multimodal vision-language model
  for segmentation of pneumothorax}.
\newblock \emph{\bibinfo{journal}{arXiv preprint arXiv:2303.01615}}
  (\bibinfo{year}{2023}).

\bibitem{hatamizadeh2022unetr}
\bibinfo{author}{Hatamizadeh, A.} \emph{et~al.}
\newblock \bibinfo{title}{Unetr: Transformers for 3d medical image
  segmentation}.
\newblock \emph{\bibinfo{journal}{2022 IEEE/CVF Winter Conference on
  Applications of Computer Vision (WACV)}} \bibinfo{pages}{1748--1758}
  (\bibinfo{year}{2022}).

\bibitem{xing2024segmamba}
\bibinfo{author}{Xing, Z.}, \bibinfo{author}{Ye, T.}, \bibinfo{author}{Yang,
  Y.}, \bibinfo{author}{Liu, G.} \& \bibinfo{author}{Zhu, L.}
\newblock \bibinfo{title}{Segmamba: Long-range sequential modeling mamba for 3d
  medical image segmentation} (\bibinfo{year}{2024}).
\newblock \eprint{2401.13560}.

\bibitem{radford2021learning}
\bibinfo{author}{Radford, A.} \emph{et~al.}
\newblock \bibinfo{title}{Learning transferable visual models from natural
  language supervision}.
\newblock In \emph{\bibinfo{booktitle}{International conference on machine
  learning}}, \bibinfo{pages}{8748--8763} (\bibinfo{organization}{PMLR},
  \bibinfo{year}{2021}).

\bibitem{hu2021lora}
\bibinfo{author}{Hu, E.~J.} \emph{et~al.}
\newblock \bibinfo{title}{Lora: Low-rank adaptation of large language models}
  (\bibinfo{year}{2021}).
\newblock \eprint{2106.09685}.

\bibitem{shen2017deep}
\bibinfo{author}{Shen, D.}, \bibinfo{author}{Wu, G.} \& \bibinfo{author}{Suk,
  H.-I.}
\newblock \bibinfo{title}{Deep learning in medical image analysis}.
\newblock \emph{\bibinfo{journal}{Annual review of biomedical engineering}}
  \textbf{\bibinfo{volume}{19}}, \bibinfo{pages}{221--248}
  (\bibinfo{year}{2017}).

\bibitem{de2018clinically}
\bibinfo{author}{De~Fauw, J.} \emph{et~al.}
\newblock \bibinfo{title}{Clinically applicable deep learning for diagnosis and
  referral in retinal disease}.
\newblock \emph{\bibinfo{journal}{Nature medicine}}
  \textbf{\bibinfo{volume}{24}}, \bibinfo{pages}{1342--1350}
  (\bibinfo{year}{2018}).

\bibitem{rajpurkar2017chexnet}
\bibinfo{author}{Rajpurkar, P.} \emph{et~al.}
\newblock \bibinfo{title}{Chexnet: Radiologist-level pneumonia detection on
  chest x-rays with deep learning}.
\newblock \emph{\bibinfo{journal}{arXiv preprint arXiv:1711.05225}}
  (\bibinfo{year}{2017}).

\bibitem{choi2019machine}
\bibinfo{author}{Choi, B.~G.} \emph{et~al.}
\newblock \bibinfo{title}{Machine learning for the prediction of new-onset
  diabetes mellitus during 5-year follow-up in non-diabetic patients with
  cardiovascular risks}.
\newblock \emph{\bibinfo{journal}{Yonsei medical journal}}
  \textbf{\bibinfo{volume}{60}}, \bibinfo{pages}{191--199}
  (\bibinfo{year}{2019}).

\bibitem{yoo2013osteoporosis}
\bibinfo{author}{Yoo, T.~K.} \emph{et~al.}
\newblock \bibinfo{title}{Osteoporosis risk prediction for bone mineral density
  assessment of postmenopausal women using machine learning}.
\newblock \emph{\bibinfo{journal}{Yonsei medical journal}}
  \textbf{\bibinfo{volume}{54}}, \bibinfo{pages}{1321--1330}
  (\bibinfo{year}{2013}).

\bibitem{hosny2018artificial}
\bibinfo{author}{Hosny, A.}, \bibinfo{author}{Parmar, C.},
  \bibinfo{author}{Quackenbush, J.}, \bibinfo{author}{Schwartz, L.~H.} \&
  \bibinfo{author}{Aerts, H.~J.}
\newblock \bibinfo{title}{Artificial intelligence in radiology}.
\newblock \emph{\bibinfo{journal}{Nature Reviews Cancer}}
  \textbf{\bibinfo{volume}{18}}, \bibinfo{pages}{500--510}
  (\bibinfo{year}{2018}).

\bibitem{tiu2022expert}
\bibinfo{author}{Tiu, E.} \emph{et~al.}
\newblock \bibinfo{title}{Expert-level detection of pathologies from
  unannotated chest x-ray images via self-supervised learning}.
\newblock \emph{\bibinfo{journal}{Nature Biomedical Engineering}}
  \textbf{\bibinfo{volume}{6}}, \bibinfo{pages}{1399--1406}
  (\bibinfo{year}{2022}).

\bibitem{moon2022multi}
\bibinfo{author}{Moon, J.~H.}, \bibinfo{author}{Lee, H.},
  \bibinfo{author}{Shin, W.}, \bibinfo{author}{Kim, Y.-H.} \&
  \bibinfo{author}{Choi, E.}
\newblock \bibinfo{title}{Multi-modal understanding and generation for medical
  images and text via vision-language pre-training}.
\newblock \emph{\bibinfo{journal}{IEEE Journal of Biomedical and Health
  Informatics}} \textbf{\bibinfo{volume}{26}}, \bibinfo{pages}{6070–6080}
  (\bibinfo{year}{2022}).
\newblock \urlprefix\url{http://dx.doi.org/10.1109/JBHI.2022.3207502}.

\bibitem{huang2023kiut}
\bibinfo{author}{Huang, Z.}, \bibinfo{author}{Zhang, X.} \&
  \bibinfo{author}{Zhang, S.}
\newblock \bibinfo{title}{Kiut: Knowledge-injected u-transformer for radiology
  report generation}.
\newblock In \emph{\bibinfo{booktitle}{Proceedings of the IEEE/CVF Conference
  on Computer Vision and Pattern Recognition}}, \bibinfo{pages}{19809--19818}
  (\bibinfo{year}{2023}).

\bibitem{hosny2022clinical}
\bibinfo{author}{Hosny, A.} \emph{et~al.}
\newblock \bibinfo{title}{Clinical validation of deep learning algorithms for
  radiotherapy targeting of non-small-cell lung cancer: an observational
  study}.
\newblock \emph{\bibinfo{journal}{The Lancet Digital Health}}
  \textbf{\bibinfo{volume}{4}}, \bibinfo{pages}{e657--e666}
  (\bibinfo{year}{2022}).

\bibitem{cciccek20163d}
\bibinfo{author}{{\c{C}}i{\c{c}}ek, {\"O}.}, \bibinfo{author}{Abdulkadir, A.},
  \bibinfo{author}{Lienkamp, S.~S.}, \bibinfo{author}{Brox, T.} \&
  \bibinfo{author}{Ronneberger, O.}
\newblock \bibinfo{title}{3d u-net: learning dense volumetric segmentation from
  sparse annotation}.
\newblock In \emph{\bibinfo{booktitle}{Medical Image Computing and
  Computer-Assisted Intervention--MICCAI 2016: 19th International Conference,
  Athens, Greece, October 17-21, 2016, Proceedings, Part II 19}},
  \bibinfo{pages}{424--432} (\bibinfo{organization}{Springer},
  \bibinfo{year}{2016}).

\bibitem{KingBa15}
\bibinfo{author}{Kingma, D.} \& \bibinfo{author}{Ba, J.}
\newblock \bibinfo{title}{Adam: A method for stochastic optimization}.
\newblock In \emph{\bibinfo{booktitle}{International Conference on Learning
  Representations (ICLR)}} (\bibinfo{address}{San Diega, CA, USA},
  \bibinfo{year}{2015}).

\bibitem{paszke2019pytorch}
\bibinfo{author}{Paszke, A.} \emph{et~al.}
\newblock \bibinfo{title}{Pytorch: An imperative style, high-performance deep
  learning library}.
\newblock \emph{\bibinfo{journal}{Advances in neural information processing
  systems}} \textbf{\bibinfo{volume}{32}} (\bibinfo{year}{2019}).

\bibitem{xu2023open}
\bibinfo{author}{Xu, J.} \emph{et~al.}
\newblock \bibinfo{title}{Open-vocabulary panoptic segmentation with
  text-to-image diffusion models}.
\newblock In \emph{\bibinfo{booktitle}{Proceedings of the IEEE/CVF Conference
  on Computer Vision and Pattern Recognition}}, \bibinfo{pages}{2955--2966}
  (\bibinfo{year}{2023}).

\bibitem{crum2006generalized}
\bibinfo{author}{Crum, W.~R.}, \bibinfo{author}{Camara, O.} \&
  \bibinfo{author}{Hill, D.~L.}
\newblock \bibinfo{title}{Generalized overlap measures for evaluation and
  validation in medical image analysis}.
\newblock \emph{\bibinfo{journal}{IEEE transactions on medical imaging}}
  \textbf{\bibinfo{volume}{25}}, \bibinfo{pages}{1451--1461}
  (\bibinfo{year}{2006}).

\bibitem{tvseg_2024_12792278}
\bibinfo{author}{Oh, Y.} \emph{et~al.}
\newblock \bibinfo{title}{Llm-driven multimodal target volume contouring in
  radiation oncology} (\bibinfo{year}{2024}).
\newblock \urlprefix\url{https://doi.org/10.5281/zenodo.12792278}.

\end{thebibliography}

\section*{Acknowledgement} This research was supported by Basic Science Research Program through the NRF funded by the Ministry of Education under Grant RS-2023-00242164 to S.P., and also supported by the National Research Foundation of Korea(NRF) grant funded by the Korea government(MSIT) (RS-2024-00336454), (RS-2023-00262527) to J.C.Y., (No. 2022R1A2C2008623) to J.S.K., and (RS-2024-00345854) to Y.O.

\section*{Author Contributions}
Y.O. designed the study, extended the code, conducted all experiments, analyzed data, and contributed to manuscript preparation. S.P. conceptualized the study, gathered and labeled the data, analyzed data, and also contributed to manuscript preparation. H.K.B., {Y.C. and I.J.L. were responsible for data collection and manuscript preparation.} J.S.K. and J.C.Y. provided supervision throughout the project, from conception to discussion, and assisted in preparing the manuscript.

\section*{Competing Interests}
J.S.K. is a shareholder and employee of Oncosoft Inc, which may benefit from the research results presented in this paper. This potential conflict of interest has been disclosed and managed according to institutional policies.

\clearpage
\newpage

\section*{Tables}

\begin{table*}[!h]
    \centering
    \caption{ {Comparison of 3D CTV delineation performance for breast cancer patients.}} 
    
    \resizebox{1\linewidth}{!}{
    \begin{tabular}{llccccccc}
    
    \toprule
    
        &  & \multicolumn{3}{c}{{Vision-only AI}} & \multicolumn{4}{c}{{Multimodal AI}}  \\
        \cmidrule(l){3-5} \cmidrule(l){6-9}

          \multirow{1}{*}{{Dataset}} & \multirow{1}{*}{{Metric}} & 3D ResUNet \cite{cciccek20163d}  & {3D SegMamba \cite{xing2024segmamba}}  & {3D UNETR \cite{hatamizadeh2022unetr}} & \add{HIPIE}\cite{wang2024hierarchical} & {LISA}\cite{lai2023lisa} & {ConTEXTualNet$^{\dagger}$} \cite{huemann2023contextual} & {LLMSeg (Ours)} \\ 
       
        \toprule

          \multirow{3}{*}{\shortstack[l]{{Internal Test}\\{(N=307)}}} & \shortstack[c]{{Dice $\uparrow$}\\{}}  &  \shortstack[c]{{0.807}\\{(0.788-0.825)}} &  \shortstack[c]{{0.699}\\{(0.679-0.718)}} &  \shortstack[c]{{0.592}\\{(0.576-0.606)}} &  \shortstack[c]{{0.743}\\{(0.732-0.754)}} &  \shortstack[c]{{0.746}\\{(0.731-0.760)}}  &    \shortstack[c]{{0.819}\\{(0.800-0.835)}} &  \shortstack[c]{{0.829}\\{(0.809-0.845)}}  \\ 
          
         & \shortstack[c]{{IoU $\uparrow$}\\{}}  &  \shortstack[c]{{0.698}\\{(0.677-0.718)}} &  \shortstack[c]{{0.559}\\{(0.538-0.580)}} &  \shortstack[c]{{0.433}\\{(0.418-0.447)}} &      \shortstack[c]{{0.600}\\{(0.587-0.613)}} & \shortstack[c]{{0.608}\\{(0.591-0.624)}}  &      \shortstack[c]{{0.715}\\{(0.695-0.733)}}   &  \shortstack[c]{{0.730}\\{(0.709-0.748)}}  \\ 
         
         & \shortstack[c]{{HD-95 $\downarrow$}\\{}} &  \shortstack[c]{{6.674}\\{(5.891-7.452)}} &  \shortstack[c]{{14.857}\\{(14.258-15.483)}} &  \shortstack[c]{{18.408}\\{(17.930-18.918)}} &  \shortstack[c]{{3.479}\\{(3.139-3.850)}} &  \shortstack[c]{{4.437}\\{(3.915-4.995)}}   &   \shortstack[c]{{4.540}\\{(3.806-5.273)}}    &  \shortstack[c]{{3.386}\\{(2.890-3.949)}}  \\ 

        \cmidrule(l){1-1} \cmidrule(l){2-2} \cmidrule(l){3-5} \cmidrule(l){6-9} 

          \multirow{3}{*}{\shortstack[l]{{External Test \#1}\\{(N=206)}}} & \shortstack[c]{{Dice $\uparrow$}\\{}}  &  \shortstack[c]{{0.731}\\{(0.707-0.755)}} &  \shortstack[c]{{0.555}\\{(0.523-0.587)}} &  \shortstack[c]{{0.522}\\{(0.508-0.535)}} &   \shortstack[c]{{0.736}\\{(0.719-0.751)}} &  \shortstack[c]{{0.691}\\{(0.669-0.712)}}   &     \shortstack[c]{{0.815}\\{(0.798-0.832)}}   &  \shortstack[c]{{0.822}\\{(0.805-0.836)}}  \\ 
         
         & \shortstack[c]{{IoU $\uparrow$}\\{}}  &  \shortstack[c]{{0.599}\\{(0.575-0.622)}} &  \shortstack[c]{{0.422}\\{(0.392-0.451)}} &  \shortstack[c]{{0.359}\\{(0.347-0.370)}} &   \shortstack[c]{{0.594}\\{(0.576-0.611)}} &   \shortstack[c]{{0.547}\\{(0.524-0.569)}} &     \shortstack[c]{{0.701}\\{(0.680-0.721)}} &   \shortstack[c]{{0.709}\\{(0.689-0.727)}}  \\
         
         & \shortstack[c]{{HD-95 $\downarrow$}\\{}} &  \shortstack[c]{{19.922}\\{(18.611-21.189)}} &  \shortstack[c]{{16.451}\\{(15.725-17.152)}} &  \shortstack[c]{{22.677}\\{(21.794-23.599)}} &   \shortstack[c]{{4.973}\\{(4.299-5.680)}} & \shortstack[c]{{10.859}\\{(9.814-11.926)}}  &   \shortstack[c]{{6.362}\\{(5.084-7.712)}}    &  \shortstack[c]{{4.256}\\{(3.471-5.176)}}  \\

        \cmidrule(l){1-1} \cmidrule(l){2-2} \cmidrule(l){3-5} \cmidrule(l){6-9} 

         \multirow{3}{*}{\shortstack[l]{{External Test \#2}\\{(N=204)}}} 
         & \shortstack[c]{{Dice $\uparrow$}\\{}}  &  \shortstack[c]{{0.444}\\{(0.419-0.469)}} &  \shortstack[c]{{0.638}\\{(0.619-0.658)}} &  \shortstack[c]{{0.565}\\{(0.554-0.576)}} &  \shortstack[c]{{0.617}\\{(0.593-0.640)}} & \shortstack[c]{{0.532}\\{(0.502-0.560)}}  &   \shortstack[c]{{0.826}\\{(0.809-0.840)}}     &  \shortstack[c]{{0.844}\\{(0.826-0.857)}}  \\ 
         
         & \shortstack[c]{{IoU $\uparrow$}\\{}}  &  \shortstack[c]{{0.302}\\{(0.282-0.322)}} &  \shortstack[c]{{0.484}\\{(0.464-0.507)}} &  \shortstack[c]{{0.399}\\{(0.388-0.409)}} & \shortstack[c]{{0.469}\\{(0.446-0.492)}} &  \shortstack[c]{{0.389}\\{(0.362-0.414)}}  &  \shortstack[c]{{0.715}\\{(0.697-0.732)}}   &  \shortstack[c]{{0.740}\\{(0.722-0.756)}}  \\
         
         & \shortstack[c]{{HD-95 $\downarrow$}\\{}} &  \shortstack[c]{{33.339}\\{(32.997-33.615)}} &  \shortstack[c]{{16.434}\\{(15.935-16.904)}} &  \shortstack[c]{{17.154}\\{(16.761-17.486)}} &   \shortstack[c]{{12.805}\\{(11.975-13.600)}} &  \shortstack[c]{{15.625}\\{(14.931-16.257)}}  &  \shortstack[c]{{5.179}\\{(4.250-6.160)}}      &  \shortstack[c]{{3.004}\\{(2.555-3.533)}} \\

     \bottomrule
     \multicolumn{8}{l}{{{Note.} $^{\dagger}$ modified for 3D CT segmentation.}}

    \end{tabular}
    }
    \label{tab_main}
\end{table*}

\clearpage

\begin{table*}[!h]
    \caption{{Expert evaluation of CTV delineation performance for breast cancer patients.}}
    \centering
    \resizebox{0.8\linewidth}{!}{
    \begin{tabular}{lcccccc}
    \toprule

    &  \multicolumn{6}{c}{{Expert Rubrics}} \\
     \cmidrule(l){2-6} \cmidrule(l){7-7}
    
    Dataset & \shortstack[c]{{Laterality}}  & \shortstack[c]{{Surgery Type}} & \shortstack[c]{{Volume Definition}} & \shortstack[c]{{Coverage}} & \shortstack[c]{{Integrity}} & \shortstack[c]{{Total}}  \\ 

       & (1 point) & (1 point) & (1.5 point) & (1 point) & (0.5 point) &  {(5 point)} \\ 
        \midrule

         & \multicolumn{6}{c}{{Vision-only AI}} \\ 
          \cmidrule(l){1-1}  \cmidrule(l){2-6} \cmidrule(l){7-7} 

        {\shortstack[l]{{}\\{Internal Test}\\{(N=307)}}} &  \shortstack[c]{{0.786}\\{(0.738-0.833)}} & \shortstack[c]{{0.887}\\{(0.854-0.918)}} & \shortstack[c]{{0.900}\\{(0.844-0.959)}} & \shortstack[c]{{0.478}\\{(0.436-0.518)}} & \shortstack[c]{{0.216}\\{(0.190-0.243)}} & \shortstack[c]{{3.267}\\{(3.139-3.385)}} \\

        {\shortstack[l]{{}\\{External Test \#1}\\{(N=206)}}} & \shortstack[c]{{0.344}\\{(0.279-0.412)}} & \shortstack[c]{{0.863}\\{(0.821-0.899)}} & \shortstack[c]{{0.680}\\{(0.615-0.748)}} & \shortstack[c]{{0.186}\\{(0.149-0.226)}} & \shortstack[c]{{0.124}\\{(0.093-0.154)}} & \shortstack[c]{{2.198}\\{(2.056-2.346)}} \\

        {\shortstack[l]{{}\\{External Test \#2}\\{(N=204)}}} & \shortstack[c]{{0.268}\\{(0.213-0.332)}} & \shortstack[c]{{0.828}\\{(0.782-0.874)}} & \shortstack[c]{{0.488}\\{(0.418-0.554)}} & \shortstack[c]{{0.029}\\{(0.015-0.047)}} & \shortstack[c]{{0.087}\\{(0.062-0.111)}} & \shortstack[c]{{1.700}\\{(1.567-1.832)}}  \\

        \midrule

          & \multicolumn{6}{c}{{Multimodal AI (LLMSeg)}}  \\ 
          \cmidrule(l){1-1}  \cmidrule(l){2-6} \cmidrule(l){7-7} 

         {\shortstack[l]{{}\\{Internal Test}\\{(N=307)}}} &  \shortstack[c]{{0.990}\\{(0.977-1.000)}} & \shortstack[c]{{0.987}\\{(0.975-0.995)}} & \shortstack[c]{{1.142}\\{(1.092-1.188)}} & \shortstack[c]{{0.602}\\{(0.562-0.641)}} & \shortstack[c]{{0.253}\\{(0.223-0.280)}} & \shortstack[c]{{3.973}\\{(3.887-4.059)}} \\

        {\shortstack[l]{{}\\{External Test \#1}\\{(N=206)}}} & \shortstack[c]{{0.990}\\{(0.976-1.000)}} & \shortstack[c]{{0.983}\\{(0.963-0.998)}} & \shortstack[c]{{1.174}\\{(1.105-1.243)}} & \shortstack[c]{{0.532}\\{(0.480-0.581)}} & \shortstack[c]{{0.260}\\{(0.226-0.294)}} & \shortstack[c]{{3.939}\\{(3.821-4.059)}} \\

        {\shortstack[l]{{}\\{External Test \#2}\\{(N=204)}}} &  \shortstack[c]{{0.990}\\{(0.975-1.000)}} & \shortstack[c]{{0.986}\\{(0.970-0.998)}} & \shortstack[c]{{1.173}\\{(1.111-1.233)}} & \shortstack[c]{{0.611}\\{(0.562-0.663)}} & \shortstack[c]{{0.250}\\{(0.215-0.287)}} & \shortstack[c]{{4.010}\\{(3.889-4.116)}} \\

    \bottomrule
    \end{tabular}
    }
    \label{tab_eval}
    \vspace{-0.2cm}
\end{table*}

\clearpage

\begin{table}[!h]
    \caption{ Ablation studies on network components.} 
    \centering
    \resizebox{1\linewidth}{!}{
    \begin{tabular}{llcccccc}
    \toprule
         {Components} & {Metric} & {LLMSeg (Ours)} & \multicolumn{2}{c}{{(a) Textual Module}} & \multicolumn{3}{c}{{(b) Tuning Method Ablation}} \\ 

         \cmidrule(l){1-1} \cmidrule(l){2-2} \cmidrule(l){3-3} \cmidrule(l){4-5} \cmidrule(l){6-8}
         
        {Text Module} &  & LLaMA-7B-Chat \cite{touvron2023llama} & {Numeric Category$^{\dagger}$} & CLIP ViT-B/16 \cite{radford2021learning} & \multicolumn{3}{c}{LLaMA-7B-Chat \cite{touvron2023llama}}  \\ 
        
        {Tuning Method} &  & Text prompts & - & \multicolumn{1}{c}{Text prompts} & One text prompt & LoRA \cite{hu2021lora} & No tuning  \\ 

        \toprule

         \multirow{3}{*}{\shortstack[l]{{}\\{Internal Test}\\{(N=307)}}} & {\shortstack[c]{{Dice $\uparrow$}\\{}}} &  \shortstack[c]{{0.829}\\{(0.809-0.845)}} &  \shortstack[c]{{0.821}\\{(0.805-0.836)}} &  \shortstack[c]{{0.813}\\{(0.795-0.830)}} &  \shortstack[c]{{0.822}\\{(0.803-0.839)}} &  \shortstack[c]{{0.829}\\{(0.812-0.844)}} &  \shortstack[c]{{0.819}\\{(0.799-0.835)}} \\ 
         
         & {\shortstack[c]{{IoU $\uparrow$}\\{}}} &  \shortstack[c]{{0.730}\\{(0.709-0.748)}} &  \shortstack[c]{{0.715}\\{(0.697-0.733)}} &  \shortstack[c]{{0.706}\\{(0.687-0.725)}} & \shortstack[c]{{0.721}\\{(0.700-0.739)}} &  \shortstack[c]{{0.726}\\{(0.708-0.742)}} &  \shortstack[c]{{0.715}\\{(0.695-0.732)}} \\ 
         
         & {\shortstack[c]{{HD-95 $\downarrow$}\\{}}} &  \shortstack[c]{{3.386}\\{(2.890-3.949)}} &  \shortstack[c]{{5.387}\\{(4.699-6.129)}} &  \shortstack[c]{{5.769}\\{(5.032-6.500)}} & \shortstack[c]{{4.036}\\{(3.442-4.697)}} &  \shortstack[c]{{3.923}\\{(3.352-4.543)}} &  \shortstack[c]{{4.546}\\{(3.877-5.210)}} \\ 

        \cmidrule(l){1-1} \cmidrule(l){2-2} \cmidrule(l){3-3} \cmidrule(l){4-5} \cmidrule(l){6-8}
        
         \multirow{3}{*}{\shortstack[l]{{}\\{External Test \#1}\\{(N=206)}}} & {\shortstack[c]{{Dice $\uparrow$}\\{}}} &  \shortstack[c]{{0.822}\\{(0.805-0.836)}} &  \shortstack[c]{{0.734}\\{(0.710-0.755)}} &  \shortstack[c]{{0.737}\\{(0.710-0.761)}} &  \shortstack[c]{{0.817}\\{(0.799-0.832)}} &  \shortstack[c]{{0.809}\\{(0.793-0.825)}} &   \shortstack[c]{{0.825}\\{(0.812-0.838)}} \\ 
         
         & {\shortstack[c]{{IoU $\uparrow$}\\{}}} &  \shortstack[c]{{0.709}\\{(0.689-0.727)}} &  \shortstack[c]{{0.601}\\{(0.577-0.623)}} &  \shortstack[c]{{0.609}\\{(0.583-0.632)}}  & \shortstack[c]{{0.703}\\{(0.683-0.722)}} &  \shortstack[c]{{0.692}\\{(0.673-0.711)}} &  \shortstack[c]{{0.712}\\{(0.695-0.729)}} \\ 
         
         & {\shortstack[c]{{HD-95 $\downarrow$}\\{}}} &  \shortstack[c]{{4.256}\\{(3.471-5.176)}} &  \shortstack[c]{{17.010}\\{(15.829-18.165)}} &  \shortstack[c]{{15.484}\\{(13.998-17.016)}} &   \shortstack[c]{{4.914}\\{(3.908-5.968)}} &  \shortstack[c]{{7.838}\\{(6.704-8.895)}} &  \shortstack[c]{{9.148}\\{(7.812-10.569)}}, \\ 

         \cmidrule(l){1-1} \cmidrule(l){2-2} \cmidrule(l){3-3} \cmidrule(l){4-5} \cmidrule(l){6-8}
        
        \multirow{3}{*}{\shortstack[l]{{}\\{External Test \#2}\\{(N=204)}}} & {\shortstack[c]{{Dice $\uparrow$}\\{}}} &  \shortstack[c]{{0.844}\\{(0.826-0.857)}} &  \shortstack[c]{{0.737}\\{(0.717-0.755)}} &  \shortstack[c]{{0.528}\\{(0.490-0.567)}} &  \shortstack[c]{{0.832}\\{(0.815-0.846)}} &  \shortstack[c]{{0.796}\\{(0.779-0.810)}} &  \shortstack[c]{{0.829}\\{(0.812-0.842)}}  \\ 
         & {\shortstack[c]{{IoU $\uparrow$}\\{}}} &  \shortstack[c]{{0.740}\\{(0.722-0.756)}} &  \shortstack[c]{{0.599}\\{(0.577-0.618)}} &  \shortstack[c]{{0.409}\\{(0.375-0.444)}}  &  \shortstack[c]{{0.724}\\{(0.705-0.739)}} &  \shortstack[c]{{0.674}\\{(0.654-0.690)}} &  \shortstack[c]{{0.718}\\{(0.700-0.733)}}  \\ 
         & {\shortstack[c]{{HD-95 $\downarrow$}\\{}}} &  \shortstack[c]{{3.004}\\{(2.555-3.533)}} &  \shortstack[c]{{13.703}\\{(12.706-14.724)}} &  \shortstack[c]{{15.429}\\{(14.312-16.583)}} & \shortstack[c]{{3.056}\\{(2.673-3.507)}} &  \shortstack[c]{{9.650}\\{(8.683-10.623)}} &   \shortstack[c]{{5.078}\\{(4.416-5.827)}}\\

         \bottomrule

         \multicolumn{8}{l}{{$^{\dagger}$ numeric categorization of text, e.g., ``0301",  corresponds to ``N0" for the first digit, `T3" for the second, ``mastectomy" for third, and ``right" for the last.}} \\
    \end{tabular}
    }
\label{tab_ablation}
\end{table}

\clearpage

\begin{table*}[!h]
    \centering
  \caption{{Ablation of input \add{clinical data} components for two different multimodal methods.}}
  \resizebox{0.8\linewidth}{!}{
    \begin{tabular}{lccccc}
        \toprule
          
          & \multicolumn{5}{c}{Ablated Input Text Components}  \\
         \cmidrule(l){2-6}   
          
         Dataset & No ablation & Laterality & N stage & T stage & Surgery \\ 
          \cmidrule(l){1-1}  \cmidrule(l){2-6}   

          & \multicolumn{5}{c}{{LLMSeg (Ours)}} \\
        \cmidrule(l){1-1} \cmidrule(l){2-6}  
        
        \multirow{3}{*}{\shortstack[l]{{}\\{Internal Test}\\{(N=307)}}} &  \shortstack[c]{{0.829}\\{(0.809-0.845)}} &  \shortstack[c]{{0.424}\\{(0.383-0.467)}} &  \shortstack[c]{{0.799}\\{(0.781-0.815)}} &  \shortstack[c]{{0.816}\\{(0.795-0.832)}} &  \shortstack[c]{{0.791}\\{(0.771-0.809)}}  \\ 
         &  \shortstack[c]{{0.730}\\{(0.709-0.748)}} &  \shortstack[c]{{0.345}\\{(0.309-0.382)}} &  \shortstack[c]{{0.686}\\{(0.667-0.702)}} &  \shortstack[c]{{0.712}\\{(0.690-0.730)}} &  \shortstack[c]{{0.680}\\{(0.659-0.699)}}  \\ 
         &  \shortstack[c]{{3.386}\\{(2.890-3.949)}} &  \shortstack[c]{{12.751}\\{(11.894-13.530)}} &  \shortstack[c]{{5.071}\\{(4.570-5.646)}} &  \shortstack[c]{{3.934}\\{(3.415-4.522)}} &  \shortstack[c]{{5.050}\\{(4.531-5.583)}}  \\ 
          \cmidrule(l){1-1}  \cmidrule(l){2-6}  
         
        \multirow{3}{*}{\shortstack[l]{{}\\{External Test \#1}\\{(N=206)}}} &   \shortstack[c]{{0.822}\\{(0.805-0.836)}} &  \shortstack[c]{{0.344}\\{(0.305-0.385)}} &  \shortstack[c]{{0.786}\\{(0.773-0.798)}} &  \shortstack[c]{{0.804}\\{(0.787-0.820)}} &  \shortstack[c]{{0.710}\\{(0.677-0.740)}}  \\ 
         &    \shortstack[c]{{0.709}\\{(0.689-0.727)}} &  \shortstack[c]{{0.251}\\{(0.219-0.283)}} &  \shortstack[c]{{0.656}\\{(0.641-0.671)}} &  \shortstack[c]{{0.686}\\{(0.666-0.705)}} &  \shortstack[c]{{0.590}\\{(0.559-0.619)}}  \\ 
         &  \shortstack[c]{{4.256}\\{(3.471-5.176)}} &  \shortstack[c]{{14.093}\\{(13.197-14.991)}} &  \shortstack[c]{{7.424}\\{(6.450-8.481)}} &  \shortstack[c]{{6.981}\\{(5.946-8.065)}} &  \shortstack[c]{{8.652}\\{(7.724-9.616)}}  \\  
        
         \cmidrule(l){1-1}  \cmidrule(l){2-6}  
         
        \multirow{3}{*}{\shortstack[l]{{}\\{External Test \#2}\\{(N=204)}}} &   \shortstack[c]{{0.844}\\{(0.826-0.857)}} &  \shortstack[c]{{0.390}\\{(0.340-0.434)}} &  \shortstack[c]{{0.805}\\{(0.790-0.817)}} &  \shortstack[c]{{0.833}\\{(0.816-0.846)}} &  \shortstack[c]{{0.763}\\{(0.740-0.784)}}  \\ 
         &  \shortstack[c]{{0.740}\\{(0.722-0.756)}} &  \shortstack[c]{{0.304}\\{(0.263-0.341)}} &  \shortstack[c]{{0.682}\\{(0.667-0.696)}} &  \shortstack[c]{{0.724}\\{(0.706-0.739)}} &  \shortstack[c]{{0.637}\\{(0.613-0.660)}}  \\ 
         &  \shortstack[c]{{3.004}\\{(2.555-3.533)}} &  \shortstack[c]{{11.971}\\{(11.161-12.814)}} &  \shortstack[c]{{4.618}\\{(4.244-5.029)}} &  \shortstack[c]{{3.954}\\{(3.475-4.475)}} &  \shortstack[c]{{7.246}\\{(6.571-7.951)}} \\ 

        \midrule
         & \multicolumn{5}{c}{{Numeric Category$^{\dagger}$} (Competing method)}  \\
        \cmidrule(l){1-1} \cmidrule(l){2-6}  

        \multirow{3}{*}{\shortstack[l]{{}\\{Internal Test}\\{(N=307)}}} &  \shortstack[c]{{0.821}\\{(0.805-0.836)}}  &  \shortstack[c]{{0.720}\\{(0.695-0.744)}} &  \shortstack[c]{{0.785}\\{(0.768-0.800)}} &  \shortstack[c]{{0.793}\\{(0.777-0.808)}} &  \shortstack[c]{{0.791}\\{(0.771-0.809)}}  \\ 
         &  \shortstack[c]{{0.715}\\{(0.697-0.733)}}  &  \shortstack[c]{{0.599}\\{(0.573-0.624)}} &  \shortstack[c]{{0.664}\\{(0.646-0.680)}} &  \shortstack[c]{{0.674}\\{(0.657-0.691)}} &  \shortstack[c]{{0.680}\\{(0.659-0.699)}}  \\ 
         &  \shortstack[c]{{5.387}\\{(4.699-6.129)}}  &  \shortstack[c]{{8.758}\\{(7.874-9.715)}} &  \shortstack[c]{{6.042}\\{(5.346-6.802)}} &  \shortstack[c]{{5.932}\\{(5.198-6.707)}} &  \shortstack[c]{{5.050}\\{(4.531-5.583)}}  \\ 
          \cmidrule(l){1-1}  \cmidrule(l){2-6}  
         
        \multirow{3}{*}{\shortstack[l]{{}\\{External Test \#1}\\{(N=206)}}} &  \shortstack[c]{{0.734}\\{(0.710-0.755)}}  &  \shortstack[c]{{0.712}\\{(0.690-0.734)}} &  \shortstack[c]{{0.724}\\{(0.695-0.749)}} &  \shortstack[c]{{0.716}\\{(0.694-0.738)}} &  \shortstack[c]{{0.710}\\{(0.677-0.740)}}  \\ 
         &  \shortstack[c]{{0.601}\\{(0.577-0.623)}}  &  \shortstack[c]{{0.573}\\{(0.551-0.596)}} &  \shortstack[c]{{0.595}\\{(0.567-0.620)}} &  \shortstack[c]{{0.579}\\{(0.557-0.603)}} &  \shortstack[c]{{0.590}\\{(0.559-0.619)}}  \\ 
         &  \shortstack[c]{{17.010}\\{(15.829-18.165)}}  &  \shortstack[c]{{19.446}\\{(18.102-20.765)}} &  \shortstack[c]{{18.058}\\{(16.906-19.214)}} &  \shortstack[c]{{20.579}\\{(19.284-21.768)}} &  \shortstack[c]{{8.652}\\{(7.724-9.616)}}  \\  
        
         \cmidrule(l){1-1}  \cmidrule(l){2-6}   
         
        \multirow{3}{*}{\shortstack[l]{{}\\{External Test \#2}\\{(N=204)}}} &  \shortstack[c]{{0.737}\\{(0.717-0.755)}}  &  \shortstack[c]{{0.555}\\{(0.513-0.595)}} &  \shortstack[c]{{0.668}\\{(0.638-0.694)}} &  \shortstack[c]{{0.669}\\{(0.638-0.698)}} &  \shortstack[c]{{0.700}\\{(0.681-0.718)}}  \\ 
         &  \shortstack[c]{{0.599}\\{(0.577-0.618)}}  &  \shortstack[c]{{0.436}\\{(0.401-0.469)}} &  \shortstack[c]{{0.529}\\{(0.502-0.555)}} &  \shortstack[c]{{0.538}\\{(0.507-0.567)}} &  \shortstack[c]{{0.553}\\{(0.534-0.573)}}   \\ 
         &  \shortstack[c]{{13.703}\\{(12.706-14.724)}} &  \shortstack[c]{{13.026}\\{(12.144-13.812)}} &  \shortstack[c]{{11.460}\\{(10.562-12.343)}} &  \shortstack[c]{{13.082}\\{(11.945-14.242)}} &  \shortstack[c]{{11.504}\\{(10.538-12.383)}}  \\ 

        \bottomrule
         \multicolumn{6}{l}{{$^{\dagger}$ for numeric categorization method, each digit is ablated, e.g., ``0301",  where the first 0 representing ``N0" is}} \\
         \multicolumn{6}{l}{{replaced with a character `?' to yield ``?301".}}  \\
        
    \end{tabular}
    }
  \label{tab_inputablation}
\end{table*}

\clearpage

\begin{table*}[!h]
    \caption{ {Comparison of 3D CTV delineation performance for prostate cancer patients}.} 
    \centering
    \resizebox{0.8\linewidth}{!}{
    \begin{tabular}{lcccccc}
    \toprule
         \multirow{2}{*}{{Metric}} & \multicolumn{3}{c}{{Vision-only AI}} &  \multicolumn{3}{c}{{Multimodal AI (LLMSeg)}} \\ 
         \cmidrule(l){2-4} \cmidrule(l){5-7}
         & {\shortstack[c]{{Dice $\uparrow$}\\{}}} & {\shortstack[c]{{IoU $\uparrow$}\\{}}} & {\shortstack[c]{{HD-95 $\downarrow$}\\{}}} & {\shortstack[c]{{Dice $\uparrow$}\\{}}} & {\shortstack[c]{{IoU $\uparrow$}\\{}}} & {\shortstack[c]{{HD-95 $\downarrow$}\\{}}}  \\
         \toprule
         
        {\shortstack[l]{{Internal Test}\\{(N=189)}}} & {\shortstack[c]{{0.725}\\{(0.697-0.751)}}} &  {\shortstack[c]{{0.598}\\{(0.568-0.626)}}} &  {\shortstack[c]{{3.522}\\{(3.084-4.014)}}} &  {\shortstack[c]{{0.754}\\{(0.729-0.779)}}} &  {\shortstack[c]{{0.631}\\{(0.604-0.658)}}} &  {\shortstack[c]{{3.036}\\{(2.646-3.482)}}}  \\
        
        \cmidrule(l){1-1} \cmidrule(l){2-4} \cmidrule(l){5-7}

        {\shortstack[l]{{External Test \#1}\\{(N=141)}}} & {\shortstack[c]{{0.682}\\{(0.656-0.705)}}} &  {\shortstack[c]{{0.535}\\{(0.508-0.560)}}} &  {\shortstack[c]{{3.762}\\{(3.463-4.102)}}} &  {\shortstack[c]{{0.729}\\{(0.706-0.750)}}} &  {\shortstack[c]{{0.589}\\{(0.563-0.613)}}} &  {\shortstack[c]{{3.522}\\{(3.194-3.932)}}} \\

    \bottomrule

    \end{tabular}
    }
    \label{tab_prostate}

\end{table*}

\clearpage

\begin{table*}[!h]
    \caption{{ Expert evaluation of CTV delineation performance for prostate cancer patients.}} 
    \centering
    \resizebox{0.8\linewidth}{!}{
    \begin{tabular}{lccccc}
    \toprule

     &  \multicolumn{5}{c}{{Expert Rubrics}} \\

        \cmidrule(l){2-5} \cmidrule(l){6-6}
        
        Dataset &  \shortstack[c]{{Primary Site}}  & \shortstack[c]{{Volume Definition}} & \shortstack[c]{{Coverage}} & \shortstack[c]{{Integrity}} & \shortstack[c]{{Total}}  \\ 

        & (1 point) & (1.5 point) & (1 point) & (0.5 point) & {(4 point)} \\ 

        \midrule

        & \multicolumn{5}{c}{{Vision-only AI}}  \\ 
        \cmidrule(l){1-1} \cmidrule(l){2-5} \cmidrule(l){6-6}

        {\shortstack[l]{{}\\{Internal Test}\\{(N=189)}}} &   \shortstack[c]{{0.470}\\{(0.412-0.529)}} & \shortstack[c]{{0.717}\\{(0.644-0.783)}} & \shortstack[c]{{0.313}\\{(0.262-0.361)}} & \shortstack[c]{{0.171}\\{(0.136-0.206)}} & \shortstack[c]{{1.670}\\{(1.527-1.810)}} \\

        {\shortstack[l]{{}\\{External Test \#1}\\{(N=141)}}} & \shortstack[c]{{0.266}\\{(0.212-0.320)}} & \shortstack[c]{{0.424}\\{(0.367-0.482)}} & \shortstack[c]{{0.115}\\{(0.079-0.147)}} & \shortstack[c]{{0.090}\\{(0.061-0.122)}} & \shortstack[c]{{0.895}\\{(0.781-1.007)}}  \\

        \midrule
        &  \multicolumn{5}{c}{{Multimodal AI (LLMSeg)}}  \\ 
         \cmidrule(l){1-1} \cmidrule(l){2-5} \cmidrule(l){6-6}

       {\shortstack[l]{{}\\{Internal Test}\\{(N=189)}}} &    \shortstack[c]{{0.583}\\{(0.529-0.639)}} & \shortstack[c]{{0.951}\\{(0.874-1.027)}} & \shortstack[c]{{0.379}\\{(0.326-0.428)}} & \shortstack[c]{{0.249}\\{(0.211-0.283)}} & \shortstack[c]{{2.162}\\{(2.003-2.310)}} \\

        {\shortstack[l]{{}\\{External Test \#1}\\{(N=141)}}} & \shortstack[c]{{0.578}\\{(0.507-0.640)}} & \shortstack[c]{{0.889}\\{(0.791-0.978)}} & \shortstack[c]{{0.248}\\{(0.205-0.295)}} & \shortstack[c]{{0.209}\\{(0.169-0.248)}} & \shortstack[c]{{1.923}\\{(1.752-2.097)}} \\
        
    \bottomrule
    \end{tabular}
    }
    \label{tab_eval_prostate}

\end{table*}

\clearpage
\newpage
\section*{Figure Legends}

\begin{figure*}[!h]
\begin{comment}
\centering
\includegraphics[width=\linewidth]{fig/fig_proposed.jpg}
\end{comment}
\caption{Overview of our proposed LLMSeg. (a) Illustration comparing the concept between the traditional vision-only AI and the multimodal AI in the context of radiotherapy target volume delineation. (b) Quantitative comparison of CTV contouring performance in the Dice metric. The Dice metric for each trial is presented with whiskers representing the range from minimum to maximum values. The center line indicates the median, the bounds of the box represent the interquartile range (from the lower quartile to the upper quartile), and the x mark indicates the mean. n denotes number of patients. The p-values indicate the statistically significant superiority of the proposed multimodal LLMSeg. All statistical tests were two-sided. (c) Visual assessment of each concept. Source data are provided as a Source Data file.}
\label{fig_proposed}
\end{figure*}

\begin{figure*}[!h]
\begin{comment}
\centering
\includegraphics[width=\linewidth]{fig/dice_efficiency.jpg}
\end{comment}
\caption{ {Comparison of target contouring performance based on varying training dataset sizes. (a) Quantitative comparison for all the validation sets. The Dice metric for each trial is presented as mean values (center lines) with 95th percentile of confidence intervals calculated with the non-parametric bootstrap method (shaded areas). n denotes number of patients. (b) Visual comparison for external validation \#1. Source data are provided as a Source Data file.}}
\label{fig_data_efficiency}
\end{figure*}

\begin{figure*}[!h]
\begin{comment}
\centering
\includegraphics[width=\linewidth]{fig/fig_modify_exp.jpg}
\end{comment}
\caption{Analysis of clinical data alignment for target contouring. (a) Illustration of modification of the input clinical data, given the same CT scan. Red font indicates modified input text. (b-c) Visual assessment of radiotherapy target contouring with modified input clinical data. }
\label{fig_modify_exp}
\end{figure*}

\begin{figure*}[!h]
\begin{comment}
\centering
\includegraphics[width=1\linewidth]{fig/figure_ablation.jpg}
\end{comment}
\caption{{{Qualitative comparison of different multimodal methods with omitted clinical data components. (a) Comparison with numeric category method: Case 1 (left breast, T2N1M0, post-mastectomy) and Case 2 (left breast, T2N1M0, post-breast conservation surgery) show our method (LLMSeg) accurately includes surgically treated areas and regional nodes, while the numeric category method inaccurately segments both breasts, missing clinical context.
(b) Omission experiment for tumor information: For right breast T1aN0M0 cancer, our method segments accurately without omission. Omitting T stage, N stage, or laterality causes incorrect regional node inclusion or opposite breast contours. The competing method is inaccurate regardless of omission.
(c) Omission experiment for surgery information: In left breast T1cN1M0 cancer post-mastectomy, our method without surgery information mimics breast-conserving surgery. The competing method inaccurately contours the opposite breast irrespective of surgery information.
}}}
\label{fig_ablation_omit}
\end{figure*}

\end{document}